\documentclass[preprint,5p,twocolumn,authoryear]{elsarticle}
\usepackage{lineno}
\usepackage{amssymb}
\usepackage{amsmath}
\usepackage{cite}
\usepackage{doi}
\usepackage{color, colortbl}

\usepackage[caption=false,font=footnotesize]{subfig}

\usepackage{xcolor}
\begin{document}

\begin{frontmatter}

%% Title, authors and addresses

%% use the tnoteref command within \title for footnotes;
%% use the tnotetext command for theassociated footnote;
%% use the fnref command within \author or \affiliation for footnotes;
%% use the fntext command for theassociated footnote;
%% use the corref command within \author for corresponding author footnotes;
%% use the cortext command for theassociated footnote;
%% use the ead command for the email address,
%% and the form \ead[url] for the home page:
%% \title{Title\tnoteref{label1}}
%% \tnotetext[label1]{}
%% \author{Name\corref{cor1}\fnref{label2}}
%% \ead{email address}
%% \ead[url]{home page}
%% \fntext[label2]{}
%% \cortext[cor1]{}
%% \affiliation{organization={},
%%            addressline={}, 
%%            city={},
%%            postcode={}, 
%%            state={},
%%            country={}}
%% \fntext[label3]{}

\title{Assessing the Ship Motion Prediction Capabilities of the Open-Source Model NEMOH Against Field Observations} %% Article title

%% use optional labels to link authors explicitly to addresses:
%% \author[label1,label2]{}
%% \affiliation[label1]{organization={},
%%             addressline={},
%%             city={},
%%             postcode={},
%%             state={},
%%             country={}}
%%
%% \affiliation[label2]{organization={},
%%             addressline={},
%%             city={},
%%             postcode={},
%%             state={},
%%             country={}}

% Author names
\author[1]{Tianshi Yu\corref{cor1}}
\ead{tians.yu7@gmail.com}
\author[1]{Ziyue Wang\fnref{fn1}}
\author[2]{Filippo Nelli}
\author[3]{Ying Tan}
\author[4]{Guillaume Ducrozet}
\author[1]{Alessandro Toffoli}

\cortext[cor1]{Corresponding author}
\fntext[fn1]{Ziyue Wang is now at Arup, Melbourne, 3008, Victoria, Australia.}

% Affiliations
\affiliation[1]{organization={Department of Infrastructure Engineering, The University of Melbourne},
                %addressline={Parkville}, 
                city={Parkville},
%               citysep={}, % Uncomment if no comma needed between city and postcode
                postcode={3010}, 
                state={Victoria},
                country={Australia}}

\affiliation[2]{organization={Department of Mechanical Engineering, Swinburne University of Technology},
                %addressline={Parkville}, 
                city={Hawthorn},
%               citysep={}, % Uncomment if no comma needed between city and postcode
                postcode={3122}, 
                state={Victoria},
                country={Australia}}

\affiliation[3]{organization={Department of Mechanical Engineering, The University of Melbourne},
                %addressline={Parkville}, 
                city={Parkville},
%               citysep={}, % Uncomment if no comma needed between city and postcode
                postcode={3010}, 
                state={Victoria},
                country={Australia}}

\affiliation[4]{organization={Nantes Université, École Centrale Nantes, CNRS, LHEEA, UMR 6598},
                %addressline={Parkville}, 
                %city={},
%               citysep={}, % Uncomment if no comma needed between city and postcode
                %postcode={}, 
                state={F-44000 Nantes},
                country={France}}

%% Abstract
\begin{abstract}
%% Text of abstract
Accurate ship motion prediction is critical for safe and efficient maritime operations, particularly in open ocean environments.  
This study evaluates the capability of NEMOH, an open-source potential flow boundary element solver, as an example of a ship motion prediction tool for real-world open ocean conditions. 
A linear model, known as the Response Amplitude Operator (RAO), is obtained using NEMOH, and is combined with the wave directional spectrum derived from the WaMoS-II marine radar on the research vessel  \textit{Akademik Tryoshnikov} during the Antarctic Circumnavigation Expedition (ACE) to predict ship motion responses in the frequency domain.  
Predictions of heave, pitch and roll are benchmarked against concurrent ship motion observations recorded by an onboard inertial measurement unit (IMU).  
The comparisons, based on the zeroth order moment of the ship motion spectrum, demonstrate a reliable heave prediction (Pearson correlation coefficient $r=0.89$, scatter index $\text{SI}=0.41$), a reasonable pitch prediction ($r=0.80$, $\text{SI}=0.47$), and an acceptable roll prediction ($r=0.63$, $\text{SI}=0.84$).  
More significant discrepancies for pitch and roll are identified under specific extreme sea conditions.  
The results demonstrate the capability of NEMOH, offering insights into its applicability for real-world maritime operations.  
\end{abstract}

%%Graphical abstract
%\begin{graphicalabstract}
%\includegraphics{grabs}
%\end{graphicalabstract}

%%Research highlights
%\begin{highlights}
%\item Research highlight 1
%\item Research highlight 2
%\end{highlights}

%% Keywords
\begin{keyword}
%% keywords here, in the form: keyword \sep keyword

%% PACS codes here, in the form: \PACS code \sep code

%% MSC codes here, in the form: \MSC code \sep code
%% or \MSC[2008] code \sep code (2000 is the default)
Ship motion \sep NEMOH \sep Response amplitude operator\sep Wave spectrum \sep WaMoS-II 

\end{keyword}

\end{frontmatter}

%% Add \usepackage{lineno} before \begin{document} and uncomment 
%% following line to enable line numbers 
%\linenumbers

%% main text
%%

\section{Introduction}\label{sec:intro}
Ship motions are described by six degrees of freedom (DoF): translational movements along the vertical axis (heave), longitudinal axis (surge), and transverse axis (sway), as well as rotational movements around these axes, namely yaw, roll, and pitch. Accurate motion prediction is essential for optimizing operations and minimizing the risks of the ship. It helps enhance navigation planning for unmanned ships to avoid hazardous responses~\citep{Chen2023AutoShipCN,zhang2022nonlinear}, improves control during operations such as maintaining a stationary position during cargo transfers~\citep{Kuchler2010Cargo}, and contributes to optimizing both passenger comfort and fuel consumption~\citep{Cademartori2023ShipMotionReview}.

The motion responses of ships arise from a combination of exogenous factors, such as waves and wind, and endogenous factors, including hull shape, weight distribution, propulsion system dynamics, stabilisers, and dynamic positioning systems. Predictions of ship motion require the development of models that incorporate these multifaceted influences.
These models are generally classified into three main categories \citep{Cademartori2023ShipMotionReview}: (i) numerical models, (ii) data-driven models, and (iii) hybrid models. Numerical models are derived from a priori knowledge of the system, such as the ship’s equation of motion~\citep{Connell2015ROM,Kuchler2010Cargo}. Theories in fluid mechanics with different levels of approximation and numerical complexity are also utilized in numrical models, such as the strip theory~\citep{Naaijen2018validationRAO,Naaijen2016ReducingRAO}, or the Navier-Stokes equations for fluid flow around a floating body, which are solved using computational fluid dynamics (CFD) techniques~\citep{Lavrov2017CFD,Kianejad2020MotionResponseCFD}. 
    
Data-driven models typically use historical data of forcing factors and past ship motion responses as inputs, with current or future ship motion as the output. The inputs and outputs are used to create datasets for model training. By applying advanced machine learning techniques, these models can uncover complex relationships between inputs and outputs that are difficult to capture with theoretical or empirical modeling~\citep{Nie2020SVR,Zhang2021LSTM,Li2024RollLSTM,Jiang2020AR}.
    
While numerical models are governed by physical laws, making them more reliable in predictable scenarios, their accuracy depends heavily on the precision of the input data and the level of detail in the model. Data-driven models, though able to capture complex patterns, may lack the interpretability and robustness provided by underlying physical laws. To leverage the strengths of both, hybrid models, which combine numerical and data-driven methods, offer a promising solution. By incorporating accurate numerical models with the adaptability of machine learning, hybrid models can provide more reliable and flexible predictions in dynamic and uncertain conditions~\citep{Dell2021Hybrid,Skulstad2020Hybrid}. As such, developing robust numerical models remains essential for improving ship motion predictability.

Numerical models mostly rely on potential flow theory for the calculation of wave-induced forces~\citep{Cademartori2023ShipMotionReview,Newman2018Hydrodynamics,zhang2023numerical}. Within this framework, the problem is solved in the frequency domain to achieve computational efficiency, assuming weak nonlinearities in both waves and structural responses. When combined with the ship’s equation of motion, this approach captures the three-dimensional interactions between the hull and surrounding waves, providing a more comprehensive model than strip theory, which assumes two-dimensional interactions for each segment of the ship~\citep[e.g.][]{Newman2018Hydrodynamics}. NEMOH, an open-source boundary element solver, applies this framework to compute hydrodynamic forces and predict ship motion in open ocean conditions~\citep{NEMOH,babarit2015theoretical}. The underlying assumption is that both the incident wave field and the resulting ship motion are linear processes, though quadratic (nonlinear) terms can be enabled in the latest version of NEMOH if needed \citep{Kurnia2023NEMOH}. This assumption simplifies the mathematical formulation and enables computational efficiency, making NEMOH a fast alternative to CFD models while maintaining reliable accuracy for practical applications.

NEMOH (version 3.0) has previously been validated for accuracy and efficiency in simulation environments~\citep{Kurnia2023NEMOH}, showing comparable results to established commercial solvers such as WAMIT~\citep{WAMIT}, DIODORE~\citep{DIODORE}, and HYDROSTAR~\citep{HYDROSTAR}, as well as to the open-source HAMS~\citep{sheng2022NEMOHvsHAMS}. A summary of selected benchmark comparisons is provided in \ref{appendix}. 
However, despite its success in controlled environments, it remains uncertain whether NEMOH can reliably predict ship motion under the complex and variable conditions of open ocean operations. This uncertainty arises because real-world wave conditions can vary widely, with storm conditions exhibiting pronounced nonlinear wave effects such as large amplitudes, steep fronts, and whitecaps~\citep[e.g.,][]{forristall2000wave,toffoli2007second,Toffoli2024SeaStats} that challenge the linear assumptions of potential flow theory. However, under the more common conditions characterized by weaker nonlinearities, the impact on structural responses is generally less severe~\citep[e.g.,][]{Xiao2020KurtosisShip,decorte2021use,yang2023numerical}. 
Nonetheless, even under typical ocean conditions, the combination of linear assumptions and the absence of viscosity can introduce uncertainties whose magnitude and operational significance remain unclear, particularly when models are applied with imperfect input data. Rather than focusing on inter-model comparisons, this study evaluates the performance of a linear potential-flow solver in a realistic operational context, where key inputs are often limited or uncertain. NEMOH was selected for its accessibility and suitability, enabling reproducible analysis with field observations. By identifying practical limitations associated with real-world complexities, the study moves beyond idealised benchmarks and contributes insights relevant to both model application and further development.

To assess NEMOH’s predictive performance in real-world scenarios, we draw on an extensive dataset of ship motion and concurrent incident wave conditions, collected over five months during the Antarctic Circumnavigation Expedition~\citep{Derkani2020Data}. The dataset is summarized in Section \S~\ref{sec:data}. The properties of both the wave field and ship motion are discussed to assess their linear or nonlinear behavior. The basic principles of NEMOH and its application for ship motion predictions are covered in Sections \S~\ref{sec:method}. A comparison between measured and predicted ship motion is presented in Section \S~\ref{sec:results}. Concluding remarks are provided in Section \S~\ref{sec:concl}.

\section{Sea state and ship motion data}\label{sec:data}

\subsection{The data set}

\begin{figure}[tbh!]
    \centering
    \includegraphics[scale=0.7]{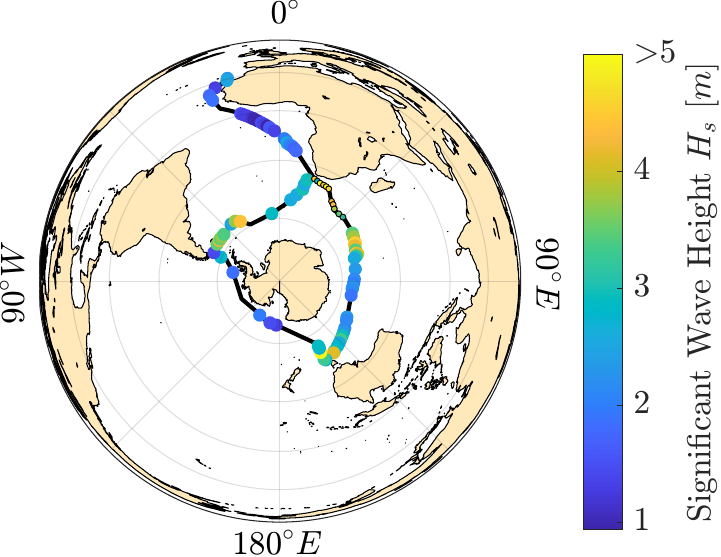}
    \caption{Antarctic Circumnavigation Expedition (ACE) route viewed from the southern pole with significant wave height ($H_s$) recorded underway. The smaller circular markers with borders represent data for calibration and the bigger ciruclar markers without a borders are the data for model assessment. While the spatial and temporal domains may appear to overlap visually, the calibration and assessment datasets are strictly non-overlapping.}
    \label{fig:route}
\end{figure}

Simultaneous measurements of sea state and ship motion were acquired during the Antarctic Circumnavigation Expedition (ACE). The expedition commenced in Cape Town (South Africa) and circumnavigated Antarctica, covering the Southern Ocean between  $34^\circ$ and $74^\circ$ South aboard the research vessel \textit{Akademik Tryoshnikov} from 20 December 2016 to 19 March 2017~\citep{Schmale2019Data, Derkani2020Data}. The full voyage also included two transit legs between the UK and South Africa. The overall route is shown in Fig. \ref{fig:route}.

The sea state characteristics were obtained using the wave and surface current monitoring system WaMoS-II. This system utilizes the ship’s X-band marine radar to capture high-resolution radar images of the surrounding ocean surface and derive the directional wave energy spectrum in the Earth-fixed reference frame at a sampling interval of 175~s~\citep{WMO1998guide,Derkani2020Data}.
To reduce the influence of short-term wave variability, a running average over a 20-minute window was applied. From each spectrum, key wave parameters—--including significant wave height, mean wave periods, spectral bandwidth, directional spreading, and mean wave direction—--were computed for the total sea state as well as for its wind sea and swell components. The significant wave height along the voyage is reported in Fig.\ref{fig:route}. It ranged from a minimum of around 1~m close to the Antarctic continent to a maximum of $\approx$6~m in the region of the Antarctic Circumpolar Current ($\approx$48-55 degrees South), noting that this value exceeds the 90th percentile normally expected in the region. The mean period varied between about 8 and 10~s (100-150~m wavelength).

Ship motion data were acquired using the inertial measurement unit (IMU) installed on the vessel, which recorded linear accelerations and rotational velocities at a sampling rate of 1 Hz. The raw measurements were processed on board to obtain three DoF in the ship’s motion: vertical translation (heave), rotation about the longitudinal axis (pitch), and rotation about the transverse axis (roll). The effects of the ship’s forward speed on the ship motion measurements were also removed during this process~\citep{fossen2011handbook}. The database was complemented with records of forward speed, ship heading, and GPS position, all acquired at the same 1 Hz sampling rate.

Time series of the ship motion were converted into power spectra using a fast Fourier transform (FFT) algorithm applied to 20-minute segments with a $50\%$ overlap. The moving window is consistent with that used to obtain the wave spectra. Quality control was applied prior to conversion by discarding segments containing outliers exceeding four times the standard deviation. Data collected within 100 km of coastlines and islands were excluded to focus the analysis on deep water and open sea conditions.

These datasets are publicly available~\citep[see details in][for ship motion and sea state data, respectively]{Alberello2020ACEMotionData,Alberello2020ACEWaMoSData}. A detailed analysis of the sea state conditions can be found in \citet[][]{Derkani2020Data}.

\subsection{Linear and nonlinear features of sea state and ship motion} \label{sec:wave_ship_nonlinear}

The wave field and the ship motion response are random processes that, under the linear assumption, follow a Gaussian distribution~\citep[cf.][]{Newman2018Hydrodynamics}. To assess the validity of Gaussian statistics or quantify deviations due to nonlinear behavior, skewness ($\lambda_3$) and kurtosis ($\kappa$) can be used~\citep[e.g.,][]{Mori2006BJDKurtosis,Onorato2009sKurtosisRange,Toffoli2024SeaStats}. The former refers to the third-order moment of the probability density function, and it measures its asymmetry. The latter is the fourth-order moment, which characterises the occurrence of extreme values or outliers in the data.

The skewness and kurtosis of ship motions are calculated directly from one-hour segments of the heave, pitch, and roll time series. 
For sea state data, provided as a directional wave spectrum, the skewness and kurtosis are approximated from the spectrum under the assumption of a narrow-banded sea state and are averaged over one hour.

The skewness of the wave field is dominated by the bound wave modes, and it is a proxy for second-order nonlinear interactions \citep[e.g.][]{Tayfun1980skewness}. It can be calculated as 
\begin{align}
    \lambda_3 = 3 \gamma^{-3} k_p M_0^{\frac{1}{2}},
    \label{eqn:skewness_bound}
\end{align}
where $M_0$ is the zero-th order moment of the wave spectrum, $\gamma=(1 + k_p^2 M_0)^{\frac{1}{2}}$ and $k_p = 4\pi^2 f_p^2 / g$ is the wave number at the spectral peak frequency $f_p$ and $g$ represents the acceleration due to gravity. The term $k_p M_0^{\frac{1}{2}}$ characterises the wave shape, with a higher value meaning sharper waves.

The kurtosis includes effects from both bound waves and quasi-resonant interactions between free modes \citep[e.g.,][]{Janssen2003BJDKurtosis}. The contribution from bound waves is calculated as \citep{Mori2011BJF2D}
\begin{align}
    \kappa_{bound} = 3 + 24k^2_pM_0.
    \label{eqn:kurtosis_bound}
\end{align}

The contribution associated with quasi-resonant interaction is \citep{janssen2019asymptotics} 
\begin{align}
    \label{eqn:kurtosis_free}
    \kappa_{free} &= 3 + \frac{\pi}{3\sqrt{3}} \frac{R_0(1-R)}{R+R_0}(\text{BFI})^2,
\end{align}
where $R_0=7.44\sqrt{3}/4\pi^3$, $R = 0.5 (\sigma_{\vartheta} / \sigma_f)^2$, $\sigma_{\vartheta}$ is the half-width (in radians) of the directional spectrum at half the maximum of the dominant spectral peak, and $\sigma_f$ is the half-width of the frequency spectrum, normalized by the peak frequency, $\text{BFI}$ is the Benjamin-Fier index, a proxy for quasi-resonant interactions \citep{Janssen2003BJDKurtosis}.
The BFI is defined as,
\begin{align}
    \text{BFI} = \frac{\sqrt{2} \epsilon}{\sigma_f}, 
    \label{eqn:BJD}
\end{align}
where $\epsilon = k_p H_s / 2$ is a form of wave steepness, and $H_s=4\sqrt{M_0}$ is the significant wave height. 

\begin{figure*}[tbh!]

    \centering
     \subfloat{
        \includegraphics[scale=0.75]{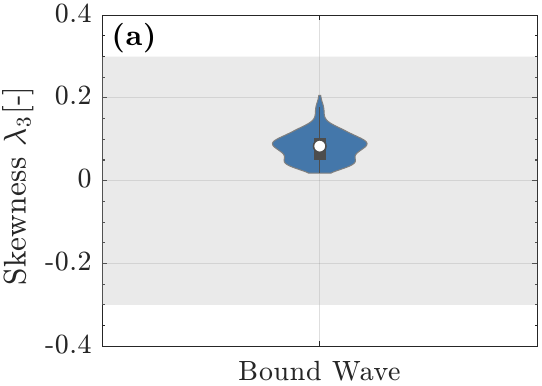}
        \label{fig:data_skewness_wave}
    }
    \hspace{10mm}
    \subfloat{
        \includegraphics[scale=0.75]{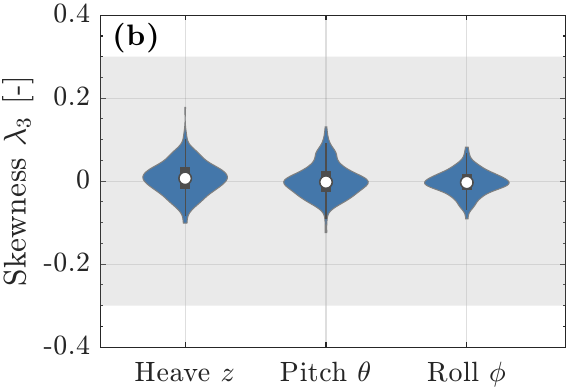}
        \label{fig:data_skewness_ship}
    }
    
    \subfloat{
        \includegraphics[scale=0.75]{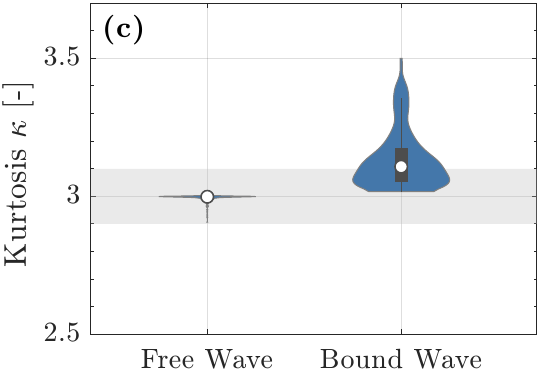}
        \label{fig:data_kurtosis_wave}
    }
    \hspace{10mm}
    \subfloat{
        \includegraphics[scale=0.75]{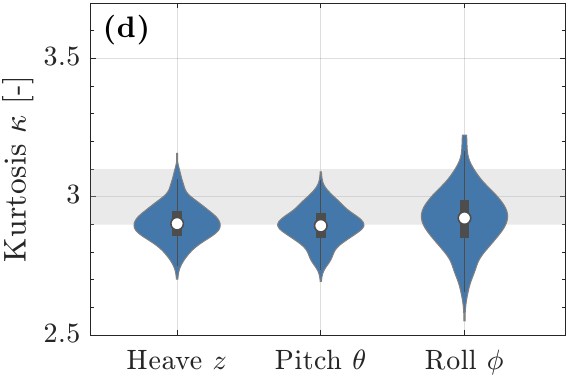}
        \label{fig:data_kurtosis_ship}
    }

    \caption{Violin plots of the skewness for (a): wave surface elevation and (b): ship heave, pitch, and roll motion, and the kurtosis for (c): wave surface and (d): ship motion. The white dots depict the medians and the black bars inside the ``violins" represent the interquartile range (IQR). The width of the ``violins" represents the density of data points at different levels. The format of the violin plot is used throughout this article. The Gaussian or weakly non-Gaussian region, ranging from -0.3 to 0.3 for skewness and 2.9 to 3.1 for kurtosis, is displayed as the grey shaded area.}
    \label{fig:data_all}
\end{figure*}

The overall distributions across the expedition are presented in Fig.~\ref{fig:data_all}, using violin plots. For a Gaussian process, skewness is zero, indicating a symmetric probability distribution. In nonlinear processes, it typically ranges between -0.3 and 0.3~\citep{Higgins1963Skewness,Onorato2009sKurtosisRange}, with a tendency toward positive values. Kurtosis is 3 for a purely Gaussian process but generally varies between 2.9 and 3.1 for weakly non-Gaussian ocean processes, although values up to 3.5 are not uncommon~\citep{Higgins1963Skewness,Onorato2009sKurtosisRange,Toffoli2024SeaStats}.

The sea states exhibit a clear nonlinear signature through weakly non-Gaussian properties. Skewness ranges between 0.05 and 0.2, with an average value of approximately 0.1 (Fig.\ref{fig:data_skewness_wave}). This non-Gaussian characteristic is further confirmed by kurtosis, which, accounting for both bound and free wave contributions, ranges between 3 and 3.5 (Fig.\ref{fig:data_kurtosis_wave}), with an average centred around 3.1. The effect of bound waves (second-order nonlinearity) dominates the kurtosis, while the influence of free waves remains marginal. 
Further evidence is provided by the wave steepness ($\epsilon$), which ranges from 0.01 to 0.13, with fewer than 5\% of cases exceeding 0.1, a threshold above which strong nonlinearity is commonly observed \citep[e.g.][]{Onorato2009sKurtosisRange,Toffoli2024SeaStats}.

For ship motion, the skewness of heave, pitch, and roll primarily oscillates around zero (Fig. \ref{fig:data_skewness_ship}), though it occasionally reaches values as high as approximately 0.1. Kurtosis remains slightly sub-Gaussian, with most values below 3 and centered around 2.9 (Fig. \ref{fig:data_kurtosis_ship}), regardless of the motion type. However, kurtosis for roll occasionally exceeds 3.1, indicating weakly non-Gaussian properties. Despite this, the overall linear response aligns with numerical simulations \citep{decorte2021use,Xiao2020KurtosisShip}, which demonstrate that nonlinearity is transferred to ship motion when quasi-resonant interactions dominate wave nonlinearity, whereas the motion response remains linear when bound waves are the dominant source of nonlinearity.

\section{Ship motion prediction using NEMOH} \label{sec:method}

To predict a ship’s motion in response to waves, NEMOH computes the Response Amplitude Operator (RAO), which characterizes the ship’s motion as a function of wave frequency and direction. The calculation is based on the ship’s equation of motion in the frequency domain, derived from potential flow theory, and solved using a boundary element method \citep[see][for details]{Kurnia2023NEMOH}. In this framework, the ship’s hull is discretized into panels, and key hydrodynamic coefficients, including added mass, radiation damping, and wave excitation forces, are determined for a range of frequencies and directions. NEMOH computes the hydrodynamic coefficients at zero speed, assuming that the primary effect of forward speed is the change in encounter frequency, which has already been accounted for in the processing of both the IMU and wave spectrum data.

The ship's equation of motion in the frequency-direction domain is given by \citep[e.g.][]{Newman2018Hydrodynamics}:
\begin{equation}
\left[\mathbf{I}+\mathbf{A}(\omega, \vartheta)\right] \ddot{\mathbf{x}} + \mathbf{B}(\omega, \vartheta) \dot{\mathbf{x}} + \mathbf{K} \mathbf{x} = \mathbf{F}(\omega, \vartheta),\label{eqn:eom}
\end{equation}
where $\mathbf{x}$ is the linear/angular displacement vector of 6 DoFs, \( \mathbf{I} \) is the inertia matrix, \( \mathbf{A}(\omega, \vartheta) \) is the frequency-dependent added mass, \( \mathbf{B}(\omega, \vartheta) \) is the frequency-dependent damping matrix, \( \mathbf{K} \) is the restoring stiffness matrix, and \( \mathbf{F}(\omega, \vartheta) \) is the excitation force vector due to wave forces. Here, \( \omega \) and \( \vartheta \) represent the wave frequency and direction, respectively. For the present work, the DoF are limited to heave, pitch, and roll to match direct observations.

The excitation forces \( \mathbf{F}(\omega, \vartheta) \) are influenced by the wave spectrum and the ship's geometry, which dictates how the hull interacts with the waves. It is considered to be proportional to the wave amplitude in linear theory.
NEMOH uses the equation of motion to compute the RAO \( (\mathbf{H}(\omega, \vartheta)) \), which establishes the relationship between the ship's displacement and unit wave amplitude (\(\mathbf{\zeta}\)) in the frequency-direction domain. The relationship is given by:
\begin{equation}
    \mathbf{x}(\omega, \vartheta) = \mathbf{H}(\omega, \vartheta) \mathbf{\zeta}(\omega,\vartheta),
\end{equation}
where
\begin{equation}
    \mathbf{H}(\omega, \vartheta) = \frac{\mathbf{F}(\omega,\vartheta)}{ \mathbf{K} - \omega^2 (\mathbf{I}+\mathbf{A}(\omega, \vartheta)) + i\omega \mathbf{B}(\omega, \vartheta)},
\end{equation}
which comes from (\ref{eqn:eom}).

Given an input incident directional wave power spectrum, \( E(\omega, \vartheta) \), measured by WaMoS-II at frequency \( \omega \) and from direction \( \vartheta \), the output ship motion power spectrum, \( R(\omega) \) for a particular DoF is computed using the following expression:
\begin{equation}
    R(\omega) = \int_{0}^{2\pi} |H(\omega, \vartheta)|^2 {E(\omega, \vartheta)} \, d\vartheta,
    \label{eqn:rao_prediction}
\end{equation}
where \( H(\omega, \vartheta) \) is the RAO associated with the DoF. This allows us to predict the ship's motion in response to waves by integrating the RAO over the wave power spectrum.

 \begin{figure}[b!]
    \centering
    \subfloat{
        \includegraphics[scale=0.7]{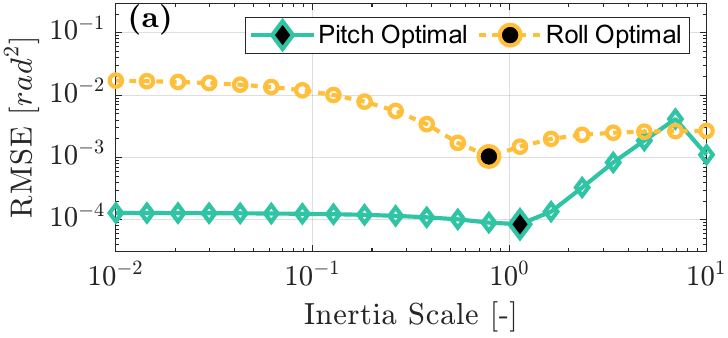}
         \label{fig:calibration_I}
    }
    
    \subfloat{
        \includegraphics[scale=0.7]{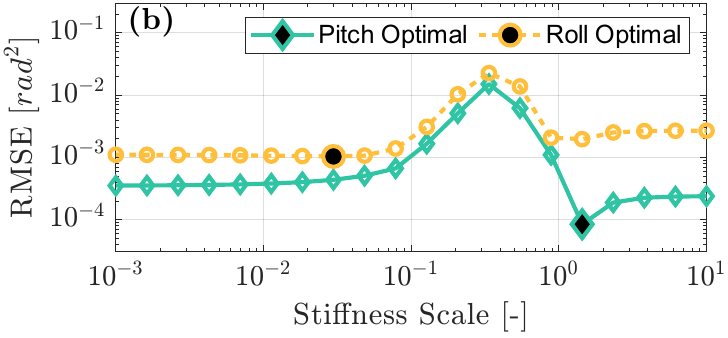}
         \label{fig:calibration_K}
    }
    \caption{Calibration results for (a) inertia and (b) stiffness in the pitch and roll axes. The curves represent the variation in prediction performance as a function of the scaling factors, with the optimal values indicated by filled black dots.}
    \label{fig:calibration}
\end{figure}

It should be noted that the damping associated with the roll motion is strongly influenced by viscous effects, making it difficult to compute accurately using potential flow methods. To address this, the simplified Ikeda method \citep{Kawahara2011Ikeda} is applied, which estimates roll damping as the sum of frictional, wave, eddy, lift, and appendage components \citep[see details for its implementation in][]{Kawahara2011Ikeda}. This approach, validated in the ITTC Recommended Procedures and Guidelines, provides a practical means of incorporating viscous effects into roll damping predictions \citep{ITTC2021Roll}.

NEMOH computes the hydrodynamic coefficients and stiffness by discretizing the ship’s hull into panels and solving the boundary integral equations for potential flow. This requires the ship geometry and wave conditions as input. To compute the RAO for the \textit{Akademik Tryoshnikov}, the ship’s key physical parameters, which define its geometry and hydrodynamic characteristics, were derived from publicly available sources. The ship has a length of 134 m, a beam of 23 m, and a nominal draft of 8.5 m \citep[see also][]{Nelli2023Satellite}. It is classified as a polar-class RS Arc7 vessel. As direct design drawings were unavailable, a similar Arc7 ice-class hull mesh was used in NEMOH under deep-water conditions. The resolution of the directional domain was 4 degrees, while that of the frequency domain was $\pi/64$ rad/s.

    \begin{figure*}[t!]
        \centering
        \subfloat{
            \includegraphics[scale =0.58]{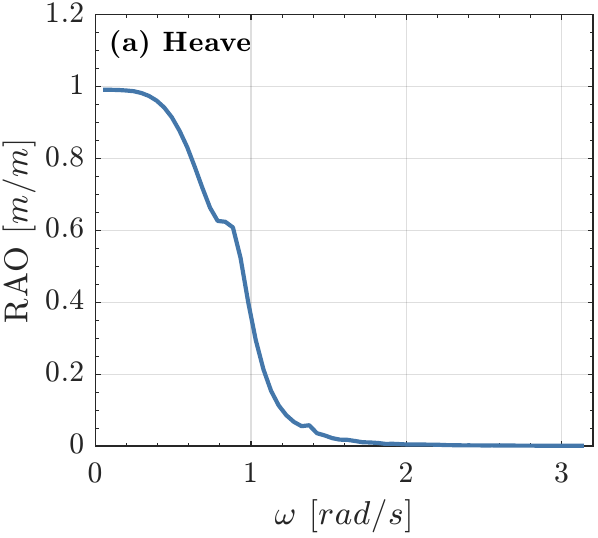}
            \label{fig:HeaveRAO}
        }
        \subfloat{
            \includegraphics[scale =0.58]{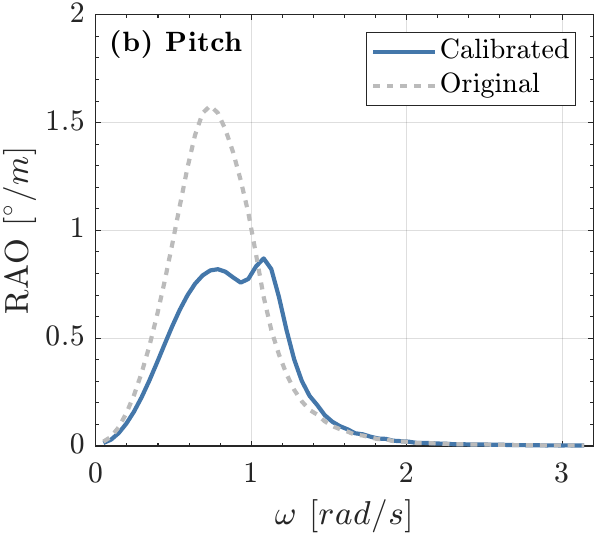}
            \label{fig:PitchRAO}
        }
        \subfloat{
            \includegraphics[scale =0.58]{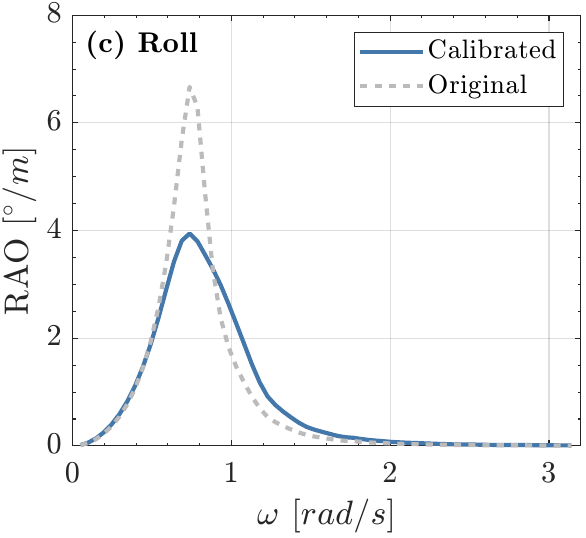}
            \label{fig:RollRAO}
        }
        \caption{RAOs of (a) heave, (b) pitch, and (c) roll used in ship motion prediction after calibration. The original RAOs for pitch and roll are also presented.}
        \label{fig:rao}
    \end{figure*}
    
The exact load configuration was unknown and, hence, calibration was necessary to refine the ship’s inertial ($\mathbf{I}$) and stiffness ($\mathbf{K}$) values for the rotational axes (pitch and roll). The coupling between DoFs is excluded, i.e., only the diagonal terms of the matrices are used for the RAO calculation, to limit the complexity of the calibration procedure. A subset of data, representing approximately $10\%$ of the overall data, was used for this purpose and is not part of any further assessment (see small circular markers with borders in Fig.\ref{fig:route}; note that the calibration and assessment datasets are strictly non-overlapping). The calibration was conducted using an iterative optimization method, where the initial $\mathbf{I}$ and $\mathbf{K}$ values were adjusted by multiplicative scaling factors to minimize errors in pitch and roll motions, as quantified by the root mean squared error of the predicted $M_0$ (see variation of the latter as a function of the adjustment factor in Fig.\ref{fig:calibration}).

The calibrated RAOs used for the prediction of heave, pitch, and roll ship motions are presented in Fig.~\ref{fig:rao}. For visualisation purposes, the frequency–direction RAO has been integrated over the directional domain ($\vartheta$).

\section{Ship motion predictions versus observations} \label{sec:results}

\subsection{Model performance}

   \begin{table}[b!]
        \caption{Ship Motion Prediction Performance}
        \centering
        \begin{tabular}{cccc}
            \hline
                & \multicolumn{3}{c}{\textbf{Metrics}}\\
             \cline{2-4}
             \textbf{DoF} & $r$ & SI & RMSE \\
             \hline
             Heave $z$ & 0.89  & 0.41  & $0.1 \ (m^2)$ \\
             Pitch $\theta$ &  0.80  & 0.47  & $4.7\times 10^{-5} (rad^2)$\\
             Roll $\phi$ &  0.63  & 0.84 & $6.6\times 10^{-4} (rad^2)$\\
             \hline
        \end{tabular}
        \label{tab:performance}
    \end{table}
    
    \begin{figure*}[tbh!]
        \centering
        \subfloat{
            \includegraphics[scale =0.5]{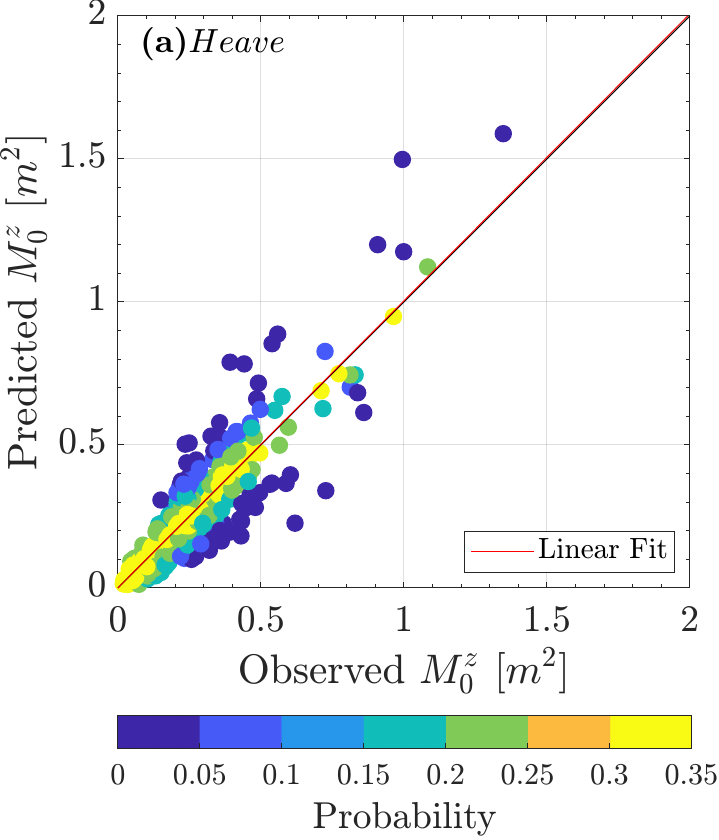}
            \label{fig:HeaveM0}
        }
        \subfloat{
            \includegraphics[scale =0.5]{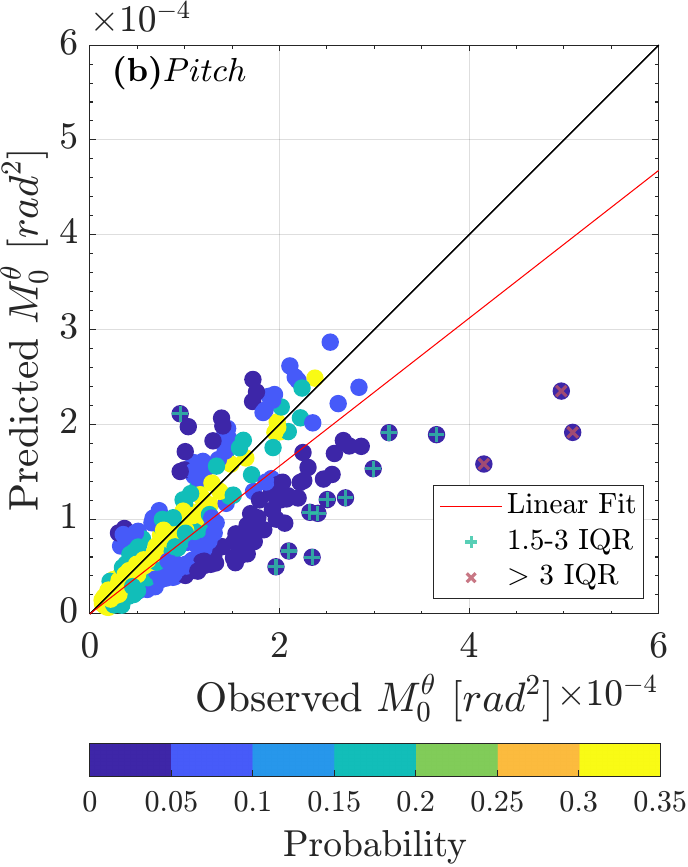}
            \label{fig:PitchM0}
        }
        \subfloat{
            \includegraphics[scale =0.5]{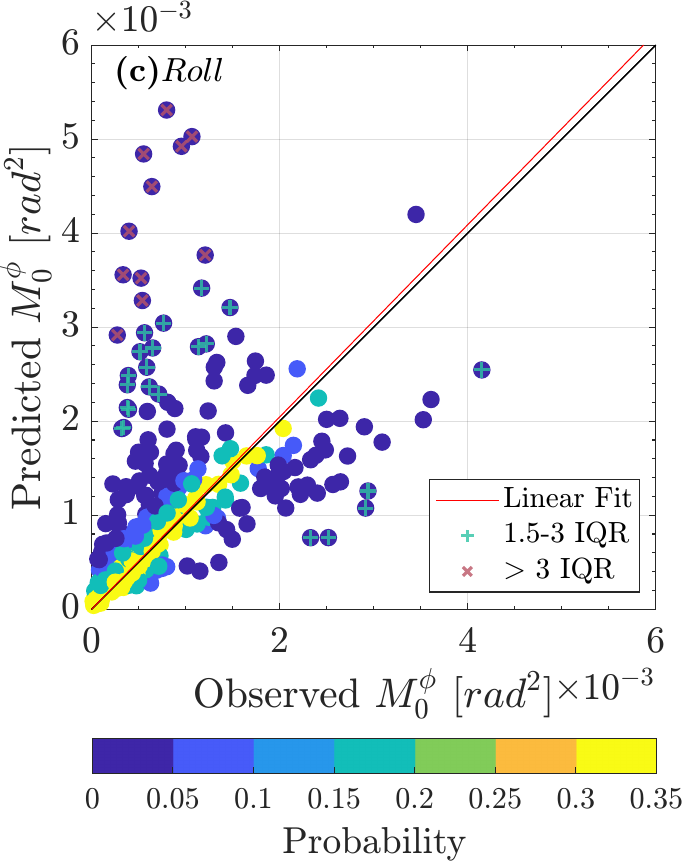}
            \label{fig:RollM0}
        }
        \caption{The scatter plot of the observed zeroth-order moment of heave, pitch and roll against the predicted values. The normal predictions are marked as filled circles. The biased predictions are marked with green + symbols, and the outlying predictions are marked with red x symbols. These are defined as being between 1.5 and 3 IQR and beyond 3 IQR, respectively, in terms of the prediction error. The color represents the occurrence probability.}
        \label{fig:comparison}
    \end{figure*}

Predictions of heave, pitch, and roll, represented by the zeroth-order moment of the concurrent spectra ($M_0^z$, $M_0^\theta$, and $M_0^\phi$, respectively), are compared with the observed values in Figs. \ref{fig:comparison}a-c. The error metrics \citep[e.g.][]{thomson2024data} are evaluated using the Pearson correlation coefficient ($r$), scatter index (SI), and root mean square error (RMSE), with the results reported in Table \ref{tab:performance}. A linear fit is included to benchmark the agreement between predictions and observations.
    
    %Heave
For heave (Fig. \ref{fig:comparison}a), the predictions align well with the observed values, with a Pearson correlation coefficient of $r = 0.89$, indicating a strong linear relationship. The linear fit closely follows the ideal correlation, reinforcing the consistency between the model and observations. Additional error metrics include a scatter index of SI = 0.41 and a root mean square error of RMSE = 0.1, reflecting a consistently low prediction error across the dataset. 

The pitch predictions exhibit a slightly more pronounced spread of values than the heave (Fig. \ref{fig:comparison}b). The error metrics are $r = 0.8$, SI = 0.47, and RMSE = $4.7 \times 10^{-5}$, suggesting a reasonable correlation and relatively small prediction error. However, the linear fit reveals a notable deviation from the ideal correlation, with predictions consistently underestimating the observations. This deviation is accentuated by outliers ($\times$ symbols in Fig. \ref{fig:comparison}b), identified as values exceeding three times the interquartile range (IQR) of the prediction error \citep{Seo2006Outlier}, as well as biased observations (+ symbols in Fig. \ref{fig:comparison}b), identified by values between 1.5 and three times the IQR. Outliers and biased values account for less than $5\%$ of the data.

Roll predictions exhibit a wider spread of values than heave and pitch (Fig. \ref{fig:comparison}c). The error metrics are $r = 0.63$, SI = 0.84, and RMSE = $6.6 \times 10^{-5}$, indicating a moderate correlation and noticeable prediction errors. These include overestimations for low to moderate roll motions due to outliers and biased data, and underestimations for more severe motions, also stemming from biased data. Despite this, the linear fit aligns with the ideal correlation. Although there is a slight trend to overestimate the observations, the overestimations and underestimations seem to offset each other.

\subsection{Sea state characteristics of outliers}
   \begin{figure*}[tb!]
        \centering

        \subfloat{
                \includegraphics[scale=0.61]{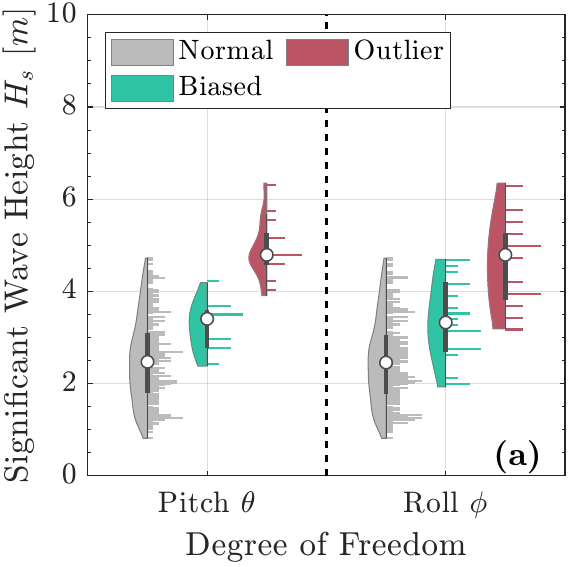}
            }
        \subfloat{
                \includegraphics[scale=0.61]{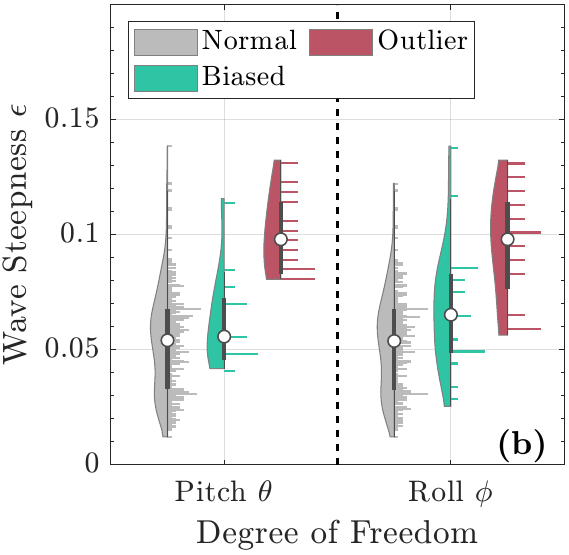}
            }
        \subfloat{
                \includegraphics[scale=0.61]{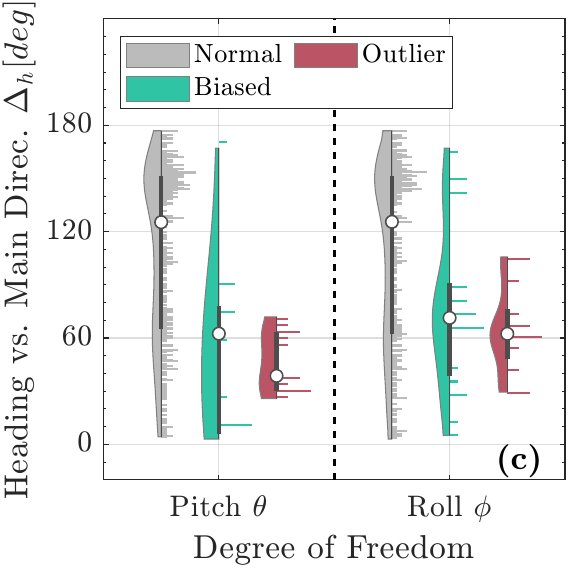}
            }
        \caption{Violin plots showing the distribution of (a) significant wave height ($H_s$), (b) wave steepness ($\epsilon$), and (c) difference between wave direction and ship’s heading ($\Delta_h$) for the normal populations, biased values, and outliers for the pitch and roll axes.}
        \label{fig:outliers}
    \end{figure*}

    \begin{figure*}[tb!]
        \centering
        \subfloat{
            \includegraphics[scale=0.7]{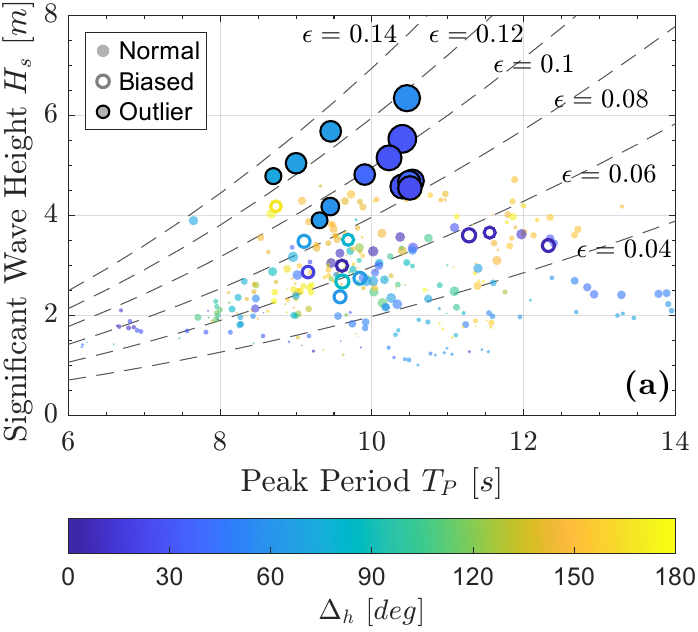}
        }
        \hspace{10pt}
        \subfloat{
            \includegraphics[scale=0.7]{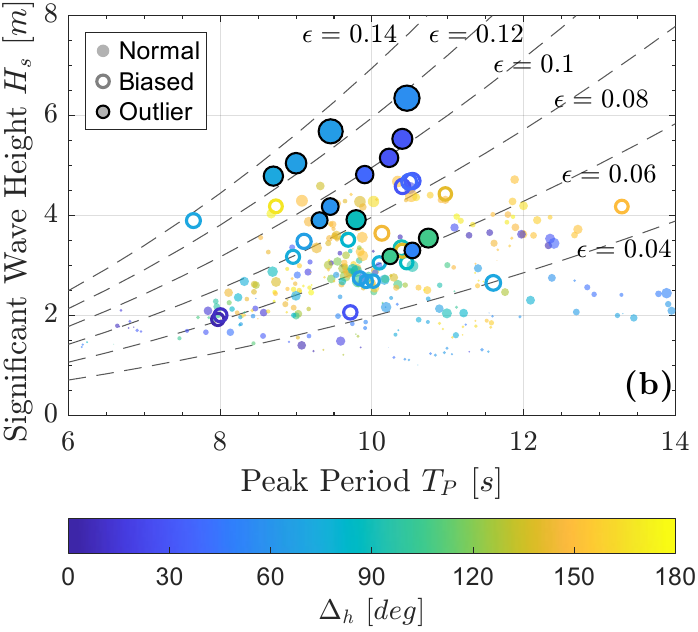}
        }
        \caption{Wave spectrum peak period $T_{p}$ vs. significant wave height $H_s$ with curves depicting wave steepness $\epsilon$, (a) pitch axis ($\theta$), (b) roll axis ($\phi$). The marker size is proportional to the prediction error, and the marker colour represents the difference between the wave direction and the ship's heading $\Delta_h$.}
        \label{fig:HsVSTp}
    \end{figure*}

Fig. \ref{fig:outliers} compares sea state parameters for the main data population, biased values, and outliers. The outliers occur in relatively severe conditions, with $H_s$ approaching or exceeding the higher values observed in the main data population: $H_s > 4$ m for pitch and $H_s > 3$ m for roll (Fig. \ref{fig:outliers}a). Wave steepness is also notable, typically exceeding $\epsilon = 0.08$ (Fig. \ref{fig:outliers}b). This steepness represents storm conditions and is often linked to ship accidents \citep{toffoli2005towards}. Pronounced steepness involves whitecaps \citep[a manifestation of localized strongly nonlinear wave interactions][]{Toffoli2024SeaStats}, which increase resistance and induce complex loads, making them challenging for potential flow models to capture accurately \citep[e.g.][]{Newman2018Hydrodynamics}. The biased values, on the other hand, show a less clear correlation with $H_s$ and $\epsilon$, as they appear to have occurred across a range of sea state conditions, from mild to severe.

Outliers also demonstrate a clear dependency on the peak wave direction relative to the ship heading ($\Delta_h$; Fig.~\ref{fig:outliers}c). For pitch, outliers predominantly occur when the sea opposes the ship’s heading ($0 \leq \Delta_h \leq 60^\circ$), while large prediction errors for roll are observed when the wave crosses the ship’s breadth. Conditions of opposing and crossing waves can induce nonlinear wave-ship interactions \citep{Newman2018Hydrodynamics}, producing resonance around the longitudinal and transverse axes, which accentuate pitch and roll motions and cannot be adequately captured by NEMOH. The concurrency of wave height, steepness, and directional effect further emphasizes coupling between heave, pitch and roll \citep{Ibrahim2010Coupling}, causing weak nonlinearity in the motion response, which aligns with the skewness observed in Figs. \ref{fig:PitchM0} and \ref{fig:RollM0}. As for $H_s$ and $\epsilon$, the biased values do not show any evident correlation with $\Delta_h$.

\begin{figure}[tb!]
        \centering
        \includegraphics[scale=0.7]{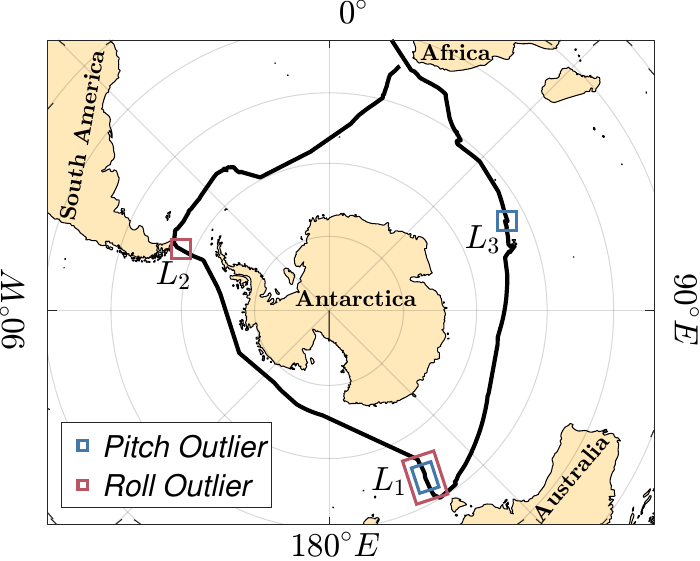}
        \caption{Geographic locations of pitch and roll prediction outliers along the ACE voyage, with $L_1$–$L_3$ denoting three distinct locations of interest (see Fig. \ref{fig:route} for the significant wave height and the full voyage route).}
        \label{fig:outlierlocations}
    \end{figure}

Fig.~\ref{fig:HsVSTp} confirms the observations from Fig.~\ref{fig:outliers}, offering a synthesised view of the correlations among the wave parameters. It is found that a combination of wave steepness, direction, and significant wave height contributes to the deterioration in prediction performance. It also reveals that most inaccuracies in the predictions occur around 10~$s$, a period corresponding to a wavelength approximately equal to the ship’s length in deep water---conditions likely to induce resonance~\citep{shin2004abscriteria}. Moreover, to provide context for the actual sailing conditions, Fig.~\ref{fig:outlierlocations} shows the geographic locations associated with the prediction outliers. Three distinct areas are identified: most outliers occur at $L_1$, where the vessel encountered a storm; a single roll outlier appears at $L_2$, corresponding to beam sea conditions; and a pitch outlier is observed at $L_3$, which also coincides with a recorded storm event\citep{Derkani2020Data}. These locations are consistent with the insights gained from the preceding analysis.

     \begin{figure*}[tb!]
        \centering

        \subfloat{
                \includegraphics[scale=0.57]{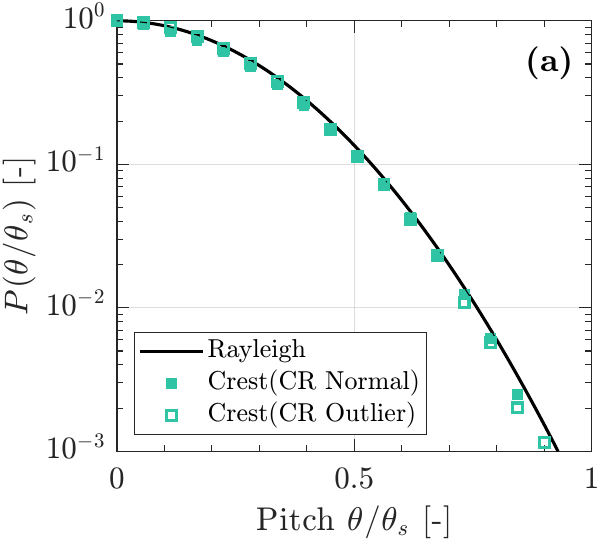}
            }
        \subfloat{
                \includegraphics[scale=0.57]{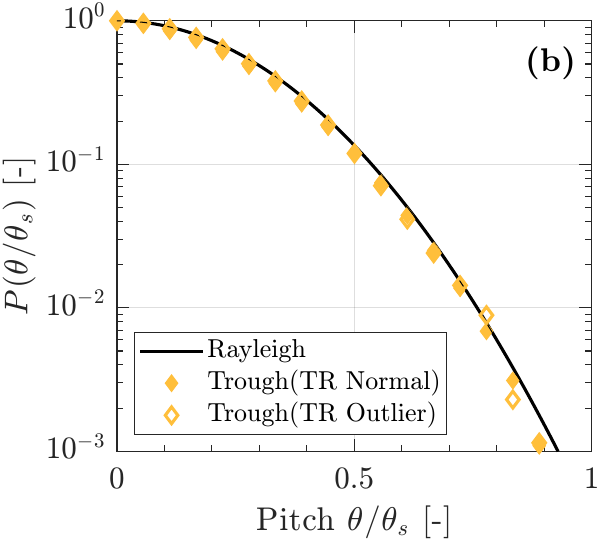}
            }
        \subfloat{
                \includegraphics[scale=0.57]{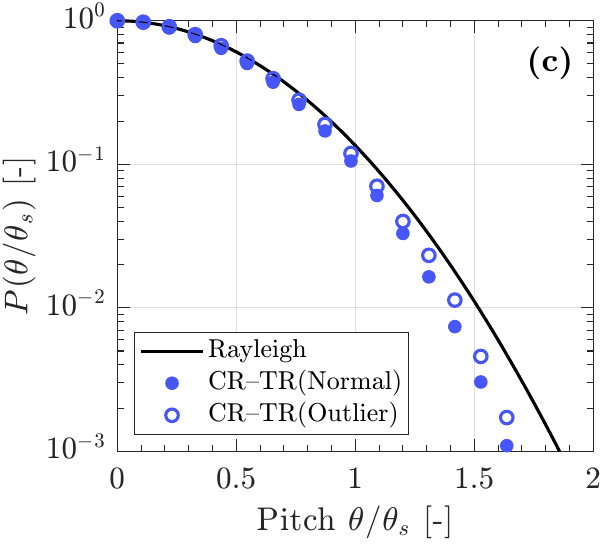}
            }
            
        \subfloat{
                \includegraphics[scale=0.57]{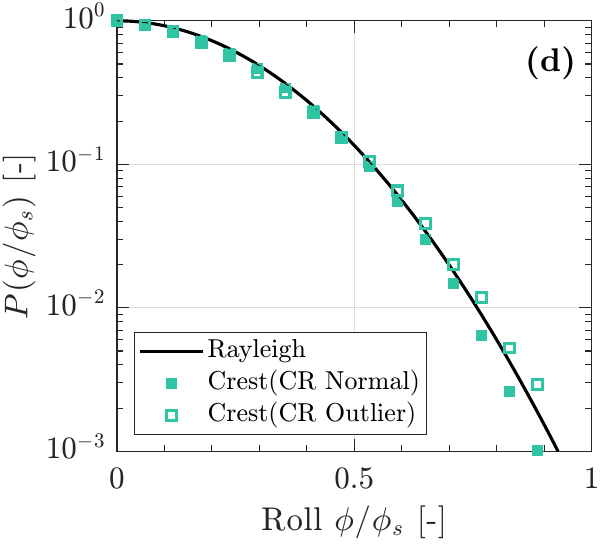}
            }
        \subfloat{
                \includegraphics[scale=0.57]{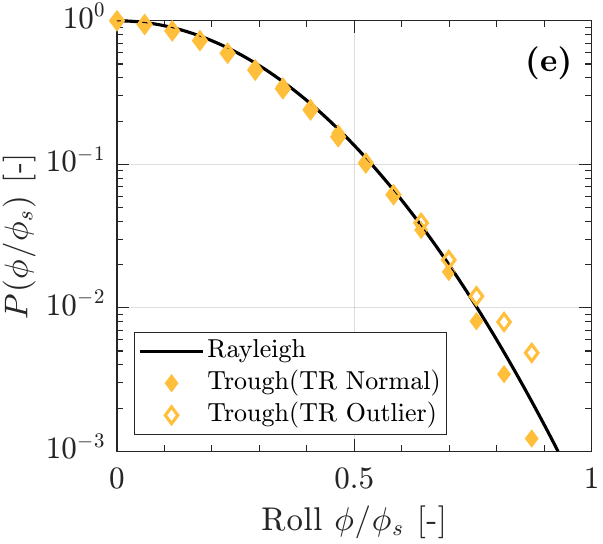}
            }
        \subfloat{
                \includegraphics[scale=0.57]{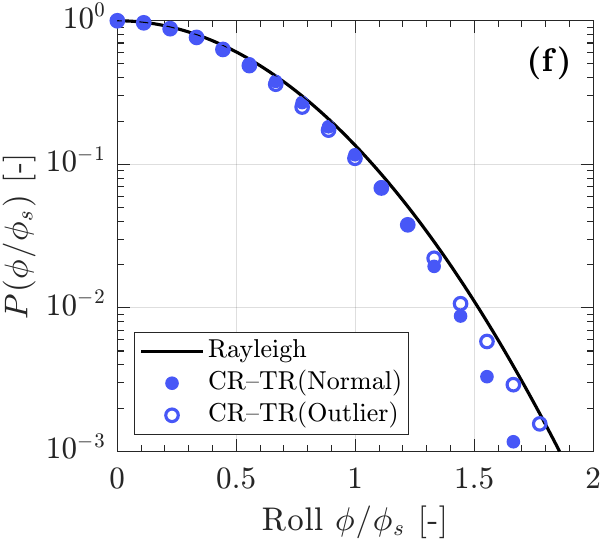}
            }
        \caption{Exceedance probability of the motion time series normalized by the $4\sqrt{M_0}$ of the motion spectrum, (a) and (d) time-series crest height of normal and outlier time series for pitch and roll, respectively; (b) and (e) time-series trough height. (c) and (f) time-series cret-to-trough height.}
        \label{fig:motion_stats}
    \end{figure*}

We note that forward speed was not fully taken into account when computing the RAO. Although the data were already corrected for encounter frequency, the forward speed has been shown to affect pitch RAO amplitude~\citep{cheng2024forwardspeed}. This effect can be significant, resulting in varying  pitch motion~\citep{cheng2024forwardspeed,Im2022forwardspeed}, which explains its underestimation in the present results. Momentum convection is another important factor influencing ship motion at non-zero forward speeds~\citep{Newman2018Hydrodynamics,Adomeit2000Convection}, but it is not accounted for in NEMOH, partially contributing to the observed bias and outlier. These limitations would require further development of NEMOH or high-fidelity CFD simulations to be addressed~\citep[e.g][]{Kianejad2020MotionResponseCFD}.

\subsection{Ship motion nonlinearity}
    
The occurrence of outliers in high, steep waves suggests the presence of nonlinear wave load effects. To examine this possibility, Fig. \ref{fig:motion_stats} presents exceedance probability statistics for peak crest, peak trough, and crest-to-trough amplitudes in the measured time series of pitch and roll motions for both the main population and outliers. The Rayleigh distribution is shown as a benchmark for linear responses, allowing deviations to be quantified.

For pitch motion (Fig. \ref{fig:motion_stats}a-c), the exceedance probabilities of the main population closely follow the Rayleigh distribution, reinforcing the linear behavior identified in Fig. \ref{fig:data_skewness_ship}. The distributions of crest and trough amplitudes for outliers do not show any significant deviation from either the main population or the Rayleigh distribution, indicating no notable nonlinear load effects. The distribution of crest-to-trough amplitudes falls slightly below the Rayleigh distribution, which reflects broadbanded sea conditions rather than non-Gaussian behavior~\citep{Newman2018Hydrodynamics,ochi1998ocean}. Relative to the main population, the crest-to-trough amplitude distribution does not show any evident deviation.

The absence of clear nonlinear effects in pitch statistics, despite its systematic under estimation (Fig. \ref{fig:PitchM0}), is herein attributed to model limitations, including simplifications in hydrodynamic forces, damping, and wave-ship interactions. Additionally, unaccounted aerodynamic effects or sensor biases contribute to uncertainty. It should also be noted that the model does not fully capture frequency-dependent resonance or transient wave impacts, which further contributes to the lower predicted values.

For roll motion, extreme values of the crest and trough amplitudes of the outliers deviate from the distribution of the main population and the Rayleigh distribution (see deviations in the upper tail of the distribution in Fig. \ref{fig:motion_stats}d–f). This departure from the linear benchmark coincides with conditions of elevated wave steepness, where nonlinear wave loads and weak nonlinear motion effects can be relevant~\citep[e.g.][]{faltinsen1990sea,faltinsen2009sloshing,maki2019nonlinearroll}. As crests become higher and troughs deepen, their combined effect increases the crest-to-trough amplitude, leading to a corresponding deviation from the distribution of the main population and the Rayleigh distribution (Fig. \ref{fig:motion_stats}e). This excess of high crest-to-trough amplitudes is consistent with the elevated kurtosis values previously reported in Fig.~\ref{fig:data_kurtosis_ship} for the roll dataset. It should be mentioned, however, that the combined influence of higher crests and deeper troughs offsets itself, leaving the asymmetry of the time series unaffected. As a result, these deviations from the linear benchmark are not detected by the skewness of the roll dataset (cf. Fig.~\ref{fig:data_skewness_ship}). 

We recall that outliers for roll motion correspond to overestimations of mild motions. In these conditions, nonlinear effects can arise from resonant or transient responses, which are not fully accounted for by the model, leading to inaccuracies in roll motion predictions. Additionally, sensitivity to specific wave characteristics introduces further uncertainty.

\section{Conclusions}\label{sec:concl}
NEMOH is an open-source boundary element solver that estimates ship motions based on linear equations, with recent updates incorporating quadratic terms. While effective in controlled environments, its reliability under complex open ocean conditions and uncertain ship parameters remains unclear.
This study evaluates NEMOH’s performance in predicting heave, pitch, and roll motion benchmarked by direct observations from the \textit{Akademik Tryoshnikov} during the Antarctic Circumnavigation Expedition. The concurrent incident wave spectra were used as input to drive the NEMOH model. The data were split between calibration and validation. Some necessary approximations were made to ship geometry and damping characteristics due to unavailability of detailed ship design data.
Sea state conditions during the voyage spanned from mild to severe conditions, with weak nonlinear properties dominated by bound wave modes (second-order effects). The weak wave nonlinearity was not transferred to the ship motion significantly, ensuring that the NEMOH operated effectively within the limit of its linear framework.
It should be noted that this study does not seek to establish comprehensive validity of the linear potential flow model, but instead aims to evaluate its performance under realistic operational conditions where input uncertainty and environmental variability are present. By doing so, it provides useful insight into the practical circumstances in which the model’s accuracy may decline, highlighting relevant sea states and factors that influence prediction reliability.

The comparison between NEMOH’s predictions and observations showed good accuracy for heave and moderate accuracy for pitch, though pitch was consistently underestimated. Roll predictions were more variable, tending to overestimate mild motions and underestimate severe ones. Large errors were linked to high wave heights and steep conditions, with opposing waves affecting pitch and crossing wave conditions increasing roll errors. The underestimations of pitch can also be attributed to the effect of forward speed on RAO amplitudes, which requires further development of NEMOH to be addressed. It is worth noting that performance limitations result from a combination of factors, including wave steepness, direction, and significant wave height, rather than a single threshold beyond which the model becomes unreliable. This analysis highlights the environmental conditions where prediction accuracy tends to decline and where NEMOH’s assumptions are increasingly challenged.

Statistics of pitch motions in sea states with large prediction errors did not reveal any evident nonlinear features. As a result, these errors were attributed to model limitations, including simplifications in motion coupling, hydrodynamic forces and damping, unaccounted aerodynamic effects, and sensor biases. In contrast, roll statistics associated with large prediction errors indicated a small but noticeable nonlinear feature, which we attributed to resonant and transient effects not accurately captured by the model. As the exact hull geometry was confidential and therefore unavailable for this study, an Arc7 ice-class hull was used in the RAO calculations, potentially introducing additional uncertainty into the predicted ship motions. Additionally, viscous damping was estimated using the simplified Ikeda method, which can introduce uncertainty in roll motion predictions under severe sea states where nonlinear effects are more pronounced \citep{ITTC2021Roll}.

Overall, while the model produced reliable estimates for heave and, to a lesser extent, pitch and roll under moderate sea states, this study clarifies its limitations in more severe conditions and under uncertain ship parameters. These findings can inform future model refinements and help users assess the applicability of linear potential flow solvers like NEMOH in operational settings.

\section*{Acknowledgments}
The ACE expedition was funded by the ACE Foundation and Ferring Pharmaceuticals. Authors acknowledge support from the Australia Research Council (LP210200927) and the WASANO project, which is funded by the French National Research Agency (ANR) as part of the ``Investisse-ments d’Aveni Programme'' (ANR-16-IDEX-0007).

%% The Appendices part is started with the command \appendix;
%% appendix sections are then done as normal sections
%\appendix
%\section{Example Appendix Section}
%\label{app1}

\appendix
\section{}
\label{appendix}
\begin{table*}[t!]
    \centering
   \caption{Summary of studies comparing NEMOH against commercial and open-source softwares}
    \begin{tabular}{p{4cm}lp{6.5cm}}
                \hline
                \rowcolor[gray]{0.9}
                 \textbf{Compared Software} & \textbf{Literature} & \textbf{Accuracy Summary} \\
                 
                 \hline
                WAMIT & \citet{nemoh_vs_wamit} & Original NEMOH shows good agreement in 1st-order coefficients.\\
                \hline
                 WAMIT, HAMS &  \citet{sheng2022NEMOHvsHAMS} & Original NEMOH closely matches WAMIT and HAMS for single body geometries. \\
                 \hline
                WAMIT, DIODORE, HYDROSTAR & \citet{Kurnia2023NEMOH} & NEMOH version 3.0 shows close results for both 1st- and 2nd-order loads.\\
                     \hline
    \end{tabular}
    \label{tab:appendix_NEMOHComparison}
\end{table*}

Table~\ref{tab:appendix_NEMOHComparison} summarises published validation results comparing NEMOH with other potential-flow solvers. These studies confirm that NEMOH produces results consistent with both commercial and open-source tools under idealised conditions.

%% For citations use: 
%%       \citet{<label>} ==> Lamport (1994)
%%       \citep{<label>} ==> (Lamport, 1994)
%%

%% If you have bib database file and want bibtex to generate the
%% bibitems, please use
%%
\bibliographystyle{elsarticle-harv} 
\bibliography{library_clean}

\begin{thebibliography}{59}
\expandafter\ifx\csname natexlab\endcsname\relax\def\natexlab#1{#1}\fi
\providecommand{\url}[1]{\texttt{#1}}
\providecommand{\href}[2]{#2}
\providecommand{\path}[1]{#1}
\providecommand{\DOIprefix}{doi:}
\providecommand{\ArXivprefix}{arXiv:}
\providecommand{\URLprefix}{URL: }
\providecommand{\Pubmedprefix}{pmid:}
\providecommand{\doi}[1]{\href{http://dx.doi.org/#1}{\path{#1}}}
\providecommand{\Pubmed}[1]{\href{pmid:#1}{\path{#1}}}
\providecommand{\bibinfo}[2]{#2}
\ifx\xfnm\relax \def\xfnm[#1]{\unskip,\space#1}\fi
%Type = Article
\bibitem[{Adomeit and Renz(2000)}]{Adomeit2000Convection}
\bibinfo{author}{Adomeit, P.}, \bibinfo{author}{Renz, U.}, \bibinfo{year}{2000}.
\newblock \bibinfo{title}{{Hydrodynamics of three-dimensional waves in laminar falling films}}.
\newblock \bibinfo{journal}{International Journal of Multiphase Flow} \bibinfo{volume}{26}, \bibinfo{pages}{1183--1208}.
%Type = Inproceedings
\bibitem[{del {\'A}guila~Ferrandis et~al.(2021)del {\'A}guila~Ferrandis, Triantafyllou, Chryssostomidis and Karniadakis}]{Dell2021Hybrid}
\bibinfo{author}{del {\'A}guila~Ferrandis, J.}, \bibinfo{author}{Triantafyllou, M.S.}, \bibinfo{author}{Chryssostomidis, C.}, \bibinfo{author}{Karniadakis, G.E.}, \bibinfo{year}{2021}.
\newblock \bibinfo{title}{{Learning functionals via LSTM neural networks for predicting vessel dynamics in extreme sea states}}, in: \bibinfo{booktitle}{Proceedings of the Royal Society A}, \bibinfo{publisher}{The Royal Society Publishing}. p. \bibinfo{pages}{20190897}.
%Type = Article
\bibitem[{Alberello et~al.(2020a)Alberello, Bennetts and Toffoli}]{Alberello2020ACEMotionData}
\bibinfo{author}{Alberello, A.}, \bibinfo{author}{Bennetts, L.}, \bibinfo{author}{Toffoli, A.}, \bibinfo{year}{2020}a.
\newblock \bibinfo{title}{{[Database] Antarctic Circumnavigation Expedition 2017: motion sensor and GPS data}} \DOIprefix\doi{10.4225/15/5A178EF0E5156}.
%Type = Article
\bibitem[{Alberello et~al.(2020b)Alberello, Bennetts and Toffoli}]{Alberello2020ACEWaMoSData}
\bibinfo{author}{Alberello, A.}, \bibinfo{author}{Bennetts, L.}, \bibinfo{author}{Toffoli, A.}, \bibinfo{year}{2020}b.
\newblock \bibinfo{title}{{[Database] Antarctic Circumnavigation Expedition 2017: WAMOS data product}} \DOIprefix\doi{10.26179/5E9D038C396F2}.
%Type = Inproceedings
\bibitem[{Babarit and Delhommeau(2015)}]{babarit2015theoretical}
\bibinfo{author}{Babarit, A.}, \bibinfo{author}{Delhommeau, G.}, \bibinfo{year}{2015}.
\newblock \bibinfo{title}{Theoretical and numerical aspects of the open source bem solver nemoh}, in: \bibinfo{booktitle}{11th European wave and tidal energy conference (EWTEC2015)}.
%Type = Article
\bibitem[{Bouws et~al.(1998)Bouws, Draper, Shearman, Laing, Feit, Mass, Eide, Francis, Carter and Battjes}]{WMO1998guide}
\bibinfo{author}{Bouws, E.}, \bibinfo{author}{Draper, L.}, \bibinfo{author}{Shearman, E.}, \bibinfo{author}{Laing, A.}, \bibinfo{author}{Feit, D.}, \bibinfo{author}{Mass, W.}, \bibinfo{author}{Eide, L.}, \bibinfo{author}{Francis, P.}, \bibinfo{author}{Carter, D.}, \bibinfo{author}{Battjes, J.}, \bibinfo{year}{1998}.
\newblock \bibinfo{title}{Guide to wave analysis and forecasting. wmo-no. 702}.
\newblock \bibinfo{journal}{World Meteorological Organization} .
%Type = Misc
\bibitem[{{Bureau Veritas}(2020)}]{HYDROSTAR}
\bibinfo{author}{{Bureau Veritas}}, \bibinfo{year}{2020}.
\newblock \bibinfo{title}{{HYDROSTAR Software v8.14}}.
%Type = Article
\bibitem[{Cademartori et~al.(2023)Cademartori, Oneto, Valdenazzi, Coraddu, Gambino and Anguita}]{Cademartori2023ShipMotionReview}
\bibinfo{author}{Cademartori, G.}, \bibinfo{author}{Oneto, L.}, \bibinfo{author}{Valdenazzi, F.}, \bibinfo{author}{Coraddu, A.}, \bibinfo{author}{Gambino, A.}, \bibinfo{author}{Anguita, D.}, \bibinfo{year}{2023}.
\newblock \bibinfo{title}{{A review on ship motions and quiescent periods prediction models}}.
\newblock \bibinfo{journal}{Ocean Engineering} \bibinfo{volume}{280}, \bibinfo{pages}{114822}.
%Type = Article
\bibitem[{Chen et~al.(2023)Chen, yip Lau, Zhang, Dulebenets, Wang and ni~Wang}]{Chen2023AutoShipCN}
\bibinfo{author}{Chen, Q.}, \bibinfo{author}{yip Lau, Y.}, \bibinfo{author}{Zhang, P.}, \bibinfo{author}{Dulebenets, M.A.}, \bibinfo{author}{Wang, N.}, \bibinfo{author}{ni~Wang, T.}, \bibinfo{year}{2023}.
\newblock \bibinfo{title}{{From concept to practicality: Unmanned vessel research in China}}.
\newblock \bibinfo{journal}{Heliyon} \bibinfo{volume}{9}, \bibinfo{pages}{e15182}.
%Type = Inproceedings
\bibitem[{Cheng et~al.(2024)Cheng, Chen and Cai}]{cheng2024forwardspeed}
\bibinfo{author}{Cheng, Z.}, \bibinfo{author}{Chen, Y.}, \bibinfo{author}{Cai, M.}, \bibinfo{year}{2024}.
\newblock \bibinfo{title}{{Effect of Forward Speed on Hydrodynamic Responses and Structural Responses of a Benchmark Vessel}}, in: \bibinfo{booktitle}{Proceedings of the International Conference on Offshore Mechanics and Arctic Engineering - OMAE}, pp. \bibinfo{pages}{1--12}.
%Type = Inproceedings
\bibitem[{Connell et~al.(2015)Connell, Rudzinsky, Brundick, Milewski, Kusters and Farquharson}]{Connell2015ROM}
\bibinfo{author}{Connell, B.S.}, \bibinfo{author}{Rudzinsky, J.P.}, \bibinfo{author}{Brundick, C.S.}, \bibinfo{author}{Milewski, W.M.}, \bibinfo{author}{Kusters, J.G.}, \bibinfo{author}{Farquharson, G.}, \bibinfo{year}{2015}.
\newblock \bibinfo{title}{{Development of an environmental and ship motion forecasting system}}, in: \bibinfo{booktitle}{International conference on offshore mechanics and arctic engineering}, \bibinfo{organization}{American Society of Mechanical Engineers}. p. \bibinfo{pages}{V011T12A058}.
%Type = Article
\bibitem[{Decorte et~al.(2021)Decorte, Toffoli, Lombaert and Monbaliu}]{decorte2021use}
\bibinfo{author}{Decorte, G.}, \bibinfo{author}{Toffoli, A.}, \bibinfo{author}{Lombaert, G.}, \bibinfo{author}{Monbaliu, J.}, \bibinfo{year}{2021}.
\newblock \bibinfo{title}{On the use of a domain decomposition strategy in obtaining response statistics in non-gaussian seas}.
\newblock \bibinfo{journal}{Fluids} \bibinfo{volume}{6}, \bibinfo{pages}{28}.
%Type = Article
\bibitem[{Derkani et~al.(2021)Derkani, Alberello, Nelli, Bennetts, Hessner, MacHutchon, Reichert, Aouf, Khan and Toffoli}]{Derkani2020Data}
\bibinfo{author}{Derkani, M.H.}, \bibinfo{author}{Alberello, A.}, \bibinfo{author}{Nelli, F.}, \bibinfo{author}{Bennetts, L.G.}, \bibinfo{author}{Hessner, K.G.}, \bibinfo{author}{MacHutchon, K.}, \bibinfo{author}{Reichert, K.}, \bibinfo{author}{Aouf, L.}, \bibinfo{author}{Khan, S.}, \bibinfo{author}{Toffoli, A.}, \bibinfo{year}{2021}.
\newblock \bibinfo{title}{{Wind, waves, and surface currents in the Southern Ocean: observations from the Antarctic Circumnavigation Expedition}}.
\newblock \bibinfo{journal}{Earth System Science Data} \bibinfo{volume}{13}, \bibinfo{pages}{1189--1209}.
%Type = Book
\bibitem[{Faltinsen(1990)}]{faltinsen1990sea}
\bibinfo{author}{Faltinsen, O.M.}, \bibinfo{year}{1990}.
\newblock \bibinfo{title}{{Sea Loads on Ships and Offshore Structures}}.
\newblock \bibinfo{publisher}{Cambridge University Press}, \bibinfo{address}{Cambridge, UK}.
%Type = Book
\bibitem[{Faltinsen and Timokha(2009)}]{faltinsen2009sloshing}
\bibinfo{author}{Faltinsen, O.M.}, \bibinfo{author}{Timokha, A.N.}, \bibinfo{year}{2009}.
\newblock \bibinfo{title}{Sloshing}.
\newblock \bibinfo{publisher}{Cambridge University Press}, \bibinfo{address}{Cambridge, UK}.
%Type = Article
\bibitem[{Forristall(2000)}]{forristall2000wave}
\bibinfo{author}{Forristall, G.Z.}, \bibinfo{year}{2000}.
\newblock \bibinfo{title}{Wave crest distributions: Observations and second-order theory}.
\newblock \bibinfo{journal}{Journal of physical oceanography} \bibinfo{volume}{30}, \bibinfo{pages}{1931--1943}.
%Type = Book
\bibitem[{Fossen(2011)}]{fossen2011handbook}
\bibinfo{author}{Fossen, T.I.}, \bibinfo{year}{2011}.
\newblock \bibinfo{title}{Handbook of marine craft hydrodynamics and motion control}.
\newblock \bibinfo{publisher}{John Wiley \& Sons}.
%Type = Article
\bibitem[{Ibrahim and Grace(2010)}]{Ibrahim2010Coupling}
\bibinfo{author}{Ibrahim, R.}, \bibinfo{author}{Grace, I.}, \bibinfo{year}{2010}.
\newblock \bibinfo{title}{{Modeling of ship roll dynamics and its coupling with heave and pitch}}.
\newblock \bibinfo{journal}{Mathematical Problems in engineering} \bibinfo{volume}{2010}, \bibinfo{pages}{934714}.
%Type = Article
\bibitem[{Im and Lee(2022)}]{Im2022forwardspeed}
\bibinfo{author}{Im, N.}, \bibinfo{author}{Lee, S.}, \bibinfo{year}{2022}.
\newblock \bibinfo{title}{{Effects of Forward Speed and Wave Height on the Seakeeping Performance of a Small Fishing Vessel}}.
\newblock \bibinfo{journal}{Journal of Marine Science and Engineering} \bibinfo{volume}{10}.
%Type = Misc
\bibitem[{ITTC(2021)}]{ITTC2021Roll}
\bibinfo{author}{ITTC}, \bibinfo{year}{2021}.
\newblock \bibinfo{title}{{ITTC Quality System Manual-Recommended Procedures and Guidelines--Estimation of Roll Damping}}.
%Type = Article
\bibitem[{Janssen(2003)}]{Janssen2003BJDKurtosis}
\bibinfo{author}{Janssen, P.A.}, \bibinfo{year}{2003}.
\newblock \bibinfo{title}{{Nonlinear four-wave interactions and freak waves}}.
\newblock \bibinfo{journal}{Journal of Physical Oceanography} \bibinfo{volume}{33}, \bibinfo{pages}{863--884}.
%Type = Article
\bibitem[{Janssen and Janssen(2019)}]{janssen2019asymptotics}
\bibinfo{author}{Janssen, P.A.}, \bibinfo{author}{Janssen, A.J.}, \bibinfo{year}{2019}.
\newblock \bibinfo{title}{Asymptotics for the long-time evolution of kurtosis of narrow-band ocean waves}.
\newblock \bibinfo{journal}{Journal of Fluid Mechanics} \bibinfo{volume}{859}, \bibinfo{pages}{790--818}.
%Type = Article
\bibitem[{Jiang et~al.(2020)Jiang, Duan, Huang, Han, Yang and Ma}]{Jiang2020AR}
\bibinfo{author}{Jiang, H.}, \bibinfo{author}{Duan, S.L.}, \bibinfo{author}{Huang, L.}, \bibinfo{author}{Han, Y.}, \bibinfo{author}{Yang, H.}, \bibinfo{author}{Ma, Q.}, \bibinfo{year}{2020}.
\newblock \bibinfo{title}{{Scale effects in AR model real-time ship motion prediction}}.
\newblock \bibinfo{journal}{Ocean Engineering} \bibinfo{volume}{203}, \bibinfo{pages}{107202}.
%Type = Article
\bibitem[{Kawahara et~al.(2011)Kawahara, Maekawa and Ikeda}]{Kawahara2011Ikeda}
\bibinfo{author}{Kawahara, Y.}, \bibinfo{author}{Maekawa, K.}, \bibinfo{author}{Ikeda, Y.}, \bibinfo{year}{2011}.
\newblock \bibinfo{title}{{A simple prediction formula of roll damping of conventional cargo ships on the basis of Ikeda's method and its limitation}}.
\newblock \bibinfo{journal}{Fluid Mechanics and its Applications} \bibinfo{volume}{97}, \bibinfo{pages}{465--486}.
%Type = Article
\bibitem[{Kianejad et~al.(2020)Kianejad, Enshaei, Duffy and Ansarifard}]{Kianejad2020MotionResponseCFD}
\bibinfo{author}{Kianejad, S.S.}, \bibinfo{author}{Enshaei, H.}, \bibinfo{author}{Duffy, J.}, \bibinfo{author}{Ansarifard, N.}, \bibinfo{year}{2020}.
\newblock \bibinfo{title}{{Investigation of a ship resonance through numerical simulation}}.
\newblock \bibinfo{journal}{Journal of Hydrodynamics} \bibinfo{volume}{32}, \bibinfo{pages}{969--983}.
%Type = Article
\bibitem[{K{\"u}chler et~al.(2010)K{\"u}chler, Mahl, Neupert, Schneider and Sawodny}]{Kuchler2010Cargo}
\bibinfo{author}{K{\"u}chler, S.}, \bibinfo{author}{Mahl, T.}, \bibinfo{author}{Neupert, J.}, \bibinfo{author}{Schneider, K.}, \bibinfo{author}{Sawodny, O.}, \bibinfo{year}{2010}.
\newblock \bibinfo{title}{{Active control for an offshore crane using prediction of the vessel’s motion}}.
\newblock \bibinfo{journal}{IEEE/ASME transactions on mechatronics} \bibinfo{volume}{16}, \bibinfo{pages}{297--309}.
%Type = Article
\bibitem[{Kurnia and Ducrozet(2023)}]{Kurnia2023NEMOH}
\bibinfo{author}{Kurnia, R.}, \bibinfo{author}{Ducrozet, G.}, \bibinfo{year}{2023}.
\newblock \bibinfo{title}{{NEMOH: Open-source boundary element solver for computation of first- and second-order hydrodynamic loads in the frequency domain}}.
\newblock \bibinfo{journal}{Computer Physics Communications} \bibinfo{volume}{292}, \bibinfo{pages}{108885}.
%Type = Article
\bibitem[{Lavrov et~al.(2017)Lavrov, Rodrigues, Gadelho and Soares}]{Lavrov2017CFD}
\bibinfo{author}{Lavrov, A.}, \bibinfo{author}{Rodrigues, J.}, \bibinfo{author}{Gadelho, J.}, \bibinfo{author}{Soares, C.G.}, \bibinfo{year}{2017}.
\newblock \bibinfo{title}{{Calculation of hydrodynamic coefficients of ship sections in roll motion using Navier-Stokes equations}}.
\newblock \bibinfo{journal}{Ocean engineering} \bibinfo{volume}{133}, \bibinfo{pages}{36--46}.
%Type = Misc
\bibitem[{{LHEEA}(2022)}]{NEMOH}
\bibinfo{author}{{LHEEA}}, \bibinfo{year}{2022}.
\newblock \bibinfo{title}{{NEMOH Software v3.0, Ecole Centrale Nantes, France}}.
\newblock \URLprefix \url{https://gitlab.com/lheea/Nemoh}.
%Type = Article
\bibitem[{Li et~al.(2024)Li, Wang, Li, Skulstad and Zhang}]{Li2024RollLSTM}
\bibinfo{author}{Li, S.}, \bibinfo{author}{Wang, T.}, \bibinfo{author}{Li, G.}, \bibinfo{author}{Skulstad, R.}, \bibinfo{author}{Zhang, H.}, \bibinfo{year}{2024}.
\newblock \bibinfo{title}{{Short-term ship roll motion prediction using the encoder–decoder Bi-LSTM with teacher forcing}}.
\newblock \bibinfo{journal}{Ocean Engineering} \bibinfo{volume}{295}, \bibinfo{pages}{116917}.
%Type = Article
\bibitem[{Longuet-Higgins(1963)}]{Higgins1963Skewness}
\bibinfo{author}{Longuet-Higgins, M.S.}, \bibinfo{year}{1963}.
\newblock \bibinfo{title}{{The effect of non-linearities on statistical distributions in the theory of sea waves}}.
\newblock \bibinfo{journal}{Journal of Fluid Mechanics} \bibinfo{volume}{17}, \bibinfo{pages}{459--480}.
%Type = Article
\bibitem[{Maki et~al.(2019)Maki, Sakai and Umeda}]{maki2019nonlinearroll}
\bibinfo{author}{Maki, A.}, \bibinfo{author}{Sakai, M.}, \bibinfo{author}{Umeda, N.}, \bibinfo{year}{2019}.
\newblock \bibinfo{title}{{Estimating a non-Gaussian probability density of the rolling motion in irregular beam seas}}.
\newblock \bibinfo{journal}{Journal of Marine Science and Technology} \bibinfo{volume}{24}, \bibinfo{pages}{1071--1077}.
%Type = Article
\bibitem[{Mori and Janssen(2006)}]{Mori2006BJDKurtosis}
\bibinfo{author}{Mori, N.}, \bibinfo{author}{Janssen, P.A.}, \bibinfo{year}{2006}.
\newblock \bibinfo{title}{{On kurtosis and occurrence probability of freak waves}}.
\newblock \bibinfo{journal}{Journal of Physical Oceanography} \bibinfo{volume}{36}, \bibinfo{pages}{1471--1483}.
%Type = Article
\bibitem[{Mori et~al.(2011)Mori, Onorato and Janssen}]{Mori2011BJF2D}
\bibinfo{author}{Mori, N.}, \bibinfo{author}{Onorato, M.}, \bibinfo{author}{Janssen, P.A.}, \bibinfo{year}{2011}.
\newblock \bibinfo{title}{{On the estimation of the kurtosis in directional sea states for freak wave forecasting}}.
\newblock \bibinfo{journal}{Journal of Physical Oceanography} \bibinfo{volume}{41}, \bibinfo{pages}{1484--1497}.
%Type = Inproceedings
\bibitem[{Naaijen et~al.(2016)Naaijen, Roozen and Huijsmans}]{Naaijen2016ReducingRAO}
\bibinfo{author}{Naaijen, P.}, \bibinfo{author}{Roozen, D.}, \bibinfo{author}{Huijsmans, R.}, \bibinfo{year}{2016}.
\newblock \bibinfo{title}{{Reducing operational risks by on-board phase resolved prediction of wave induced ship motions}}, in: \bibinfo{booktitle}{International conference on offshore mechanics and arctic engineering}, \bibinfo{organization}{American Society of Mechanical Engineers}. p. \bibinfo{pages}{V007T06A013}.
%Type = Inproceedings
\bibitem[{Naaijen et~al.(2018)Naaijen, Van~Oosten, Roozen and van't Veer}]{Naaijen2018validationRAO}
\bibinfo{author}{Naaijen, P.}, \bibinfo{author}{Van~Oosten, K.}, \bibinfo{author}{Roozen, K.}, \bibinfo{author}{van't Veer, R.}, \bibinfo{year}{2018}.
\newblock \bibinfo{title}{{Validation of a deterministic wave and ship motion prediction system}}, in: \bibinfo{booktitle}{International Conference on Offshore Mechanics and Arctic Engineering}, \bibinfo{organization}{American Society of Mechanical Engineers}. p. \bibinfo{pages}{V07BT06A032}.
%Type = Article
\bibitem[{Nelli et~al.(2023)Nelli, Derkani, Alberello and Toffoli}]{Nelli2023Satellite}
\bibinfo{author}{Nelli, F.}, \bibinfo{author}{Derkani, M.H.}, \bibinfo{author}{Alberello, A.}, \bibinfo{author}{Toffoli, A.}, \bibinfo{year}{2023}.
\newblock \bibinfo{title}{{A satellite altimetry data assimilation approach to optimise sea state estimates from vessel motion}}.
\newblock \bibinfo{journal}{Applied Ocean Research} \bibinfo{volume}{132}, \bibinfo{pages}{103479}.
%Type = Book
\bibitem[{Newman(2018)}]{Newman2018Hydrodynamics}
\bibinfo{author}{Newman, J.N.}, \bibinfo{year}{2018}.
\newblock \bibinfo{title}{Marine hydrodynamics}.
\newblock \bibinfo{publisher}{The MIT press}.
%Type = Article
\bibitem[{Nie et~al.(2020)Nie, Shen, Xu and Li}]{Nie2020SVR}
\bibinfo{author}{Nie, Z.}, \bibinfo{author}{Shen, F.}, \bibinfo{author}{Xu, D.}, \bibinfo{author}{Li, Q.}, \bibinfo{year}{2020}.
\newblock \bibinfo{title}{{An EMD-SVR model for short-term prediction of ship motion using mirror symmetry and SVR algorithms to eliminate EMD boundary effect}}.
\newblock \bibinfo{journal}{Ocean Engineering} \bibinfo{volume}{217}, \bibinfo{pages}{107927}.
%Type = Book
\bibitem[{Ochi(1998)}]{ochi1998ocean}
\bibinfo{author}{Ochi, M.K.}, \bibinfo{year}{1998}.
\newblock \bibinfo{title}{Ocean waves}.
\newblock \bibinfo{publisher}{Cambridge University Press}.
%Type = Article
\bibitem[{Onorato et~al.(2009)Onorato, Cavaleri, Fouques, Gramstad, Janssen, Monbaliu, Osborne, Pakozdi, Serio, Stansberg et~al.}]{Onorato2009sKurtosisRange}
\bibinfo{author}{Onorato, M.}, \bibinfo{author}{Cavaleri, L.}, \bibinfo{author}{Fouques, S.}, \bibinfo{author}{Gramstad, O.}, \bibinfo{author}{Janssen, P.A.}, \bibinfo{author}{Monbaliu, J.}, \bibinfo{author}{Osborne, A.R.}, \bibinfo{author}{Pakozdi, C.}, \bibinfo{author}{Serio, M.}, \bibinfo{author}{Stansberg, C.}, et~al., \bibinfo{year}{2009}.
\newblock \bibinfo{title}{{Statistical properties of mechanically generated surface gravity waves: a laboratory experiment in a three-dimensional wave basin}}.
\newblock \bibinfo{journal}{Journal of Fluid Mechanics} \bibinfo{volume}{627}, \bibinfo{pages}{235--257}.
%Type = Inproceedings
\bibitem[{Penalba et~al.(2017)Penalba, Kelly and Ringwood}]{nemoh_vs_wamit}
\bibinfo{author}{Penalba, M.}, \bibinfo{author}{Kelly, T.}, \bibinfo{author}{Ringwood, J.}, \bibinfo{year}{2017}.
\newblock \bibinfo{title}{Using nemoh for modelling wave energy converters: A comparative study with wamit}, in: \bibinfo{booktitle}{12th European Wave and Tidal Energy Conference (EWTEC)}, \bibinfo{address}{Cork}.
%Type = Misc
\bibitem[{{Principia}(2021)}]{DIODORE}
\bibinfo{author}{{Principia}}, \bibinfo{year}{2021}.
\newblock \bibinfo{title}{Diodore software v5.5.2}.
%Type = Article
\bibitem[{Schmale et~al.(2019)Schmale, Baccarini, Thurnherr, Henning, Efraim, Regayre, Bolas, Hartmann, Welti, Lehtipalo et~al.}]{Schmale2019Data}
\bibinfo{author}{Schmale, J.}, \bibinfo{author}{Baccarini, A.}, \bibinfo{author}{Thurnherr, I.}, \bibinfo{author}{Henning, S.}, \bibinfo{author}{Efraim, A.}, \bibinfo{author}{Regayre, L.}, \bibinfo{author}{Bolas, C.}, \bibinfo{author}{Hartmann, M.}, \bibinfo{author}{Welti, A.}, \bibinfo{author}{Lehtipalo, K.}, et~al., \bibinfo{year}{2019}.
\newblock \bibinfo{title}{{Overview of the Antarctic circumnavigation expedition: Study of preindustrial-like aerosols and their climate effects (ACE-SPACE)}}.
\newblock \bibinfo{journal}{Bulletin of the American Meteorological Society} \bibinfo{volume}{100}, \bibinfo{pages}{2260--2283}.
%Type = Phdthesis
\bibitem[{Seo(2006)}]{Seo2006Outlier}
\bibinfo{author}{Seo, S.}, \bibinfo{year}{2006}.
\newblock \bibinfo{title}{A review and comparison of methods for detecting outliers in univariate data sets}.
\newblock Ph.D. thesis. University of Pittsburgh.
%Type = Article
\bibitem[{Sheng et~al.(2022)Sheng, Tapoglou, Ma, Taylor, Dorrell, Parsons and Aggidis}]{sheng2022NEMOHvsHAMS}
\bibinfo{author}{Sheng, W.}, \bibinfo{author}{Tapoglou, E.}, \bibinfo{author}{Ma, X.}, \bibinfo{author}{Taylor, C.J.}, \bibinfo{author}{Dorrell, R.M.}, \bibinfo{author}{Parsons, D.R.}, \bibinfo{author}{Aggidis, G.}, \bibinfo{year}{2022}.
\newblock \bibinfo{title}{{Hydrodynamic studies of floating structures: Comparison of wave-structure interaction modelling}}.
\newblock \bibinfo{journal}{Ocean Engineering} \bibinfo{volume}{249}, \bibinfo{pages}{110878}.
%Type = Article
\bibitem[{Shin et~al.(2004)Shin, Belenky, Paulling, Weems and Lin}]{shin2004abscriteria}
\bibinfo{author}{Shin, Y.S.}, \bibinfo{author}{Belenky, V.}, \bibinfo{author}{Paulling, J.}, \bibinfo{author}{Weems, K.}, \bibinfo{author}{Lin, W.}, \bibinfo{year}{2004}.
\newblock \bibinfo{title}{Criteria for parametric roll of large containerships in longitudinal seas}.
\newblock \bibinfo{journal}{ABS Technical Papers} , \bibinfo{pages}{117--147}.
%Type = Article
\bibitem[{Skulstad et~al.(2020)Skulstad, Li, Fossen, Vik and Zhang}]{Skulstad2020Hybrid}
\bibinfo{author}{Skulstad, R.}, \bibinfo{author}{Li, G.}, \bibinfo{author}{Fossen, T.I.}, \bibinfo{author}{Vik, B.}, \bibinfo{author}{Zhang, H.}, \bibinfo{year}{2020}.
\newblock \bibinfo{title}{{A hybrid approach to motion prediction for ship docking—Integration of a neural network model into the ship dynamic model}}.
\newblock \bibinfo{journal}{IEEE Transactions on Instrumentation and Measurement} \bibinfo{volume}{70}, \bibinfo{pages}{1--11}.
%Type = Article
\bibitem[{Tayfun(1980)}]{Tayfun1980skewness}
\bibinfo{author}{Tayfun, M.A.}, \bibinfo{year}{1980}.
\newblock \bibinfo{title}{{Narrow-band nonlinear sea waves}}.
\newblock \bibinfo{journal}{Journal of Geophysical Research: Oceans} \bibinfo{volume}{85}, \bibinfo{pages}{1548--1552}.
%Type = Book
\bibitem[{Thomson and Emery(2024)}]{thomson2024data}
\bibinfo{author}{Thomson, R.E.}, \bibinfo{author}{Emery, W.J.}, \bibinfo{year}{2024}.
\newblock \bibinfo{title}{Data analysis methods in physical oceanography}.
\newblock \bibinfo{publisher}{Elsevier}.
%Type = Article
\bibitem[{Toffoli et~al.(2024)Toffoli, Alberello, Clarke, Nelli, Benetazzo, Bergamasco, Ntamba, Vichi and Onorato}]{Toffoli2024SeaStats}
\bibinfo{author}{Toffoli, A.}, \bibinfo{author}{Alberello, A.}, \bibinfo{author}{Clarke, H.}, \bibinfo{author}{Nelli, F.}, \bibinfo{author}{Benetazzo, A.}, \bibinfo{author}{Bergamasco, F.}, \bibinfo{author}{Ntamba, B.N.}, \bibinfo{author}{Vichi, M.}, \bibinfo{author}{Onorato, M.}, \bibinfo{year}{2024}.
\newblock \bibinfo{title}{{Observations of Rogue Seas in the Southern Ocean}}.
\newblock \bibinfo{journal}{Physical Review Letters} \bibinfo{volume}{132}, \bibinfo{pages}{154101}.
%Type = Article
\bibitem[{Toffoli et~al.(2005)Toffoli, Lefevre, Bitner-Gregersen and Monbaliu}]{toffoli2005towards}
\bibinfo{author}{Toffoli, A.}, \bibinfo{author}{Lefevre, J.M.}, \bibinfo{author}{Bitner-Gregersen, E.}, \bibinfo{author}{Monbaliu, J.}, \bibinfo{year}{2005}.
\newblock \bibinfo{title}{Towards the identification of warning criteria: analysis of a ship accident database}.
\newblock \bibinfo{journal}{Applied Ocean Research} \bibinfo{volume}{27}, \bibinfo{pages}{281--291}.
%Type = Article
\bibitem[{Toffoli et~al.(2007)Toffoli, Monbaliu, Onorato, Osborne, Babanin and Bitner-Gregersen}]{toffoli2007second}
\bibinfo{author}{Toffoli, A.}, \bibinfo{author}{Monbaliu, J.}, \bibinfo{author}{Onorato, M.}, \bibinfo{author}{Osborne, A.}, \bibinfo{author}{Babanin, A.}, \bibinfo{author}{Bitner-Gregersen, E.}, \bibinfo{year}{2007}.
\newblock \bibinfo{title}{Second-order theory and setup in surface gravity waves: a comparison with experimental data}.
\newblock \bibinfo{journal}{Journal of physical oceanography} \bibinfo{volume}{37}, \bibinfo{pages}{2726--2739}.
%Type = Misc
\bibitem[{{WAMIT Inc}(2021)}]{WAMIT}
\bibinfo{author}{{WAMIT Inc}}, \bibinfo{year}{2021}.
\newblock \bibinfo{title}{{WAMIT Software}}.
\newblock \URLprefix \url{https://www.wamit.com}.
%Type = Article
\bibitem[{Xiao et~al.(2020)Xiao, Zhou and Zhu}]{Xiao2020KurtosisShip}
\bibinfo{author}{Xiao, Q.}, \bibinfo{author}{Zhou, W.}, \bibinfo{author}{Zhu, R.}, \bibinfo{year}{2020}.
\newblock \bibinfo{title}{{Effects of wave-field nonlinearity on motions of ship advancing in irregular waves using HOS method}}.
\newblock \bibinfo{journal}{Ocean Engineering} \bibinfo{volume}{199}, \bibinfo{pages}{106947}.
%Type = Article
\bibitem[{Yang et~al.(2023)Yang, Zhang, Zhu, Yan and Li}]{yang2023numerical}
\bibinfo{author}{Yang, Y.}, \bibinfo{author}{Zhang, F.}, \bibinfo{author}{Zhu, R.}, \bibinfo{author}{Yan, Q.}, \bibinfo{author}{Li, Y.}, \bibinfo{year}{2023}.
\newblock \bibinfo{title}{Numerical investigations on nonlinear motions of ships sailing in large-amplitude waves based on hybrid hobem}.
\newblock \bibinfo{journal}{Ocean Engineering} \bibinfo{volume}{272}, \bibinfo{pages}{113933}.
%Type = Article
\bibitem[{Zhang et~al.(2021)Zhang, Zheng and Liu}]{Zhang2021LSTM}
\bibinfo{author}{Zhang, T.}, \bibinfo{author}{Zheng, X.Q.}, \bibinfo{author}{Liu, M.X.}, \bibinfo{year}{2021}.
\newblock \bibinfo{title}{{Multiscale attention-based LSTM for ship motion prediction}}.
\newblock \bibinfo{journal}{Ocean Engineering} \bibinfo{volume}{230}, \bibinfo{pages}{109066}.
%Type = Article
\bibitem[{Zhang et~al.(2023)Zhang, Song and Beck}]{zhang2023numerical}
\bibinfo{author}{Zhang, X.}, \bibinfo{author}{Song, X.}, \bibinfo{author}{Beck, R.F.}, \bibinfo{year}{2023}.
\newblock \bibinfo{title}{Numerical investigations on seakeeping and added resistance in head waves based on nonlinear potential flow methods}.
\newblock \bibinfo{journal}{Ocean Engineering} \bibinfo{volume}{276}, \bibinfo{pages}{114043}.
%Type = Article
\bibitem[{Zhang et~al.(2022)Zhang, Zou, Yang and Zhang}]{zhang2022nonlinear}
\bibinfo{author}{Zhang, Z.}, \bibinfo{author}{Zou, Z.}, \bibinfo{author}{Yang, J.}, \bibinfo{author}{Zhang, X.}, \bibinfo{year}{2022}.
\newblock \bibinfo{title}{Nonlinear hydrodynamic model based course control and roll stabilization by taking rudder and propeller actions}.
\newblock \bibinfo{journal}{Ocean Engineering} \bibinfo{volume}{263}, \bibinfo{pages}{112377}.

\end{thebibliography}

%% else use the following coding to input the bibitems directly in the
%% TeX file.

%% Refer following link for more details about bibliography and citations.
%% https://en.wikibooks.org/wiki/LaTeX/Bibliography_Management

\end{document}